\documentclass[aps, prd, superscriptaddress, nofootinbib, preprintnumbers]{revtex4-2}
\usepackage[colorlinks=true]{hyperref}
\usepackage{xcolor}
\hypersetup{colorlinks=true, citecolor=blue, urlcolor=blue, linkcolor=blue}
\usepackage{amsmath,amssymb}
\usepackage[final]{graphicx}
\usepackage{subcaption}
\usepackage[belowskip=0.5pt,aboveskip=0.5pt]{caption}
\usepackage{float}
\usepackage{adjustbox}
\usepackage{cancel,soul,ulem}

\usepackage{todonotes}

\begin{document}

\title{Probing the chiral and $U(1)$ axial symmetry restoration via meson susceptibilities in holographic QCD}

\author{Hiwa A. Ahmed}
\thanks{{\tt hiwa.ahmed@chu.edu.iq}}
\affiliation{Charmo Centre for Research, Training and Consultancy, Charmo University, 46023, Chamchamal, Sulaymaniyah, Iraq}
\author{Danning Li}\thanks{{\tt lidanning@jnu.edu.cn}}
\affiliation{Department of Physics and Siyuan Laboratory, Jinan University, Guangzhou 510632, P.R. China}
\author{Mamiya Kawaguchi}\thanks{{\tt mamiya@aust.edu.cn}} 
\affiliation{Centre for Fundamental Physics, School of Mechanics and Photoelectric Physics, Anhui University of Science and Technology, Huainan, Anhui 232001, People’s Republic of China}
\author{Mei Huang}\thanks{{\tt huangmei@ucas.ac.cn}}
\affiliation{School of Nuclear Science and Technology, University of Chinese Academy of Sciences, Beijing 100049, China}

\date{\today}

\begin{abstract}

We investigate the restoration patterns of chiral and $U(1)$ axial symmetries at finite temperature using a soft-wall holographic QCD model. The study employs two distinct parameter sets (Case I and Case II), both calibrated to reproduce a pseudocritical temperature $T_{\rm pc} \sim 155$ MeV and the physical pion mass. The temperature dependence of the light and strange quark condensates confirms a smooth chiral crossover transition, with pseudocritical temperatures of $T_{\rm pc}=0.157$ GeV and $T_{\rm pc}=0.154$ GeV for Cases I and II, respectively. The screening masses of chiral partner mesons ($\pi$-$\sigma$ and $\eta$-$a_0$) become degenerate near $T_{\rm pc}$, providing a clear signature of chiral symmetry restoration. Analysis of the corresponding meson susceptibilities further supports this conclusion. However, the effective restoration of the $U(1)_A$ symmetry—probed via the susceptibility difference $\chi_\pi - \chi_{a_0}$—occurs more slowly than that of chiral symmetry. This indicates a separation between the chiral and axial symmetry restoration scales within the present holographic framework.  
The temperature-dependent topological susceptibility $\chi_{\rm top}^{1/4}$ is also computed, showing a sharp drop near $T_{\rm pc}$ and a subsequent slight decrease. While the model qualitatively captures established features of the chiral transition, the results highlight a limitation in the quantitative description of the $U(1)$ axial anomaly compared to LQCD in our work.

\end{abstract}


\maketitle

\section{Introduction}

In the study of quantum chromodynamics (QCD), there is significant interest in understanding the restoration of chiral symmetry and the $U(1)$ axial ($U(1)_A$) anomaly under extreme conditions such as high temperature or density. In the massless quark limit (chiral limit), the QCD Lagrangian exhibits chiral symmetry; however, this symmetry is spontaneously broken in the vacuum, giving rise to the hadron mass spectrum and other key properties \cite{Hatsuda:1994pi}. The $U(1)_A$ symmetry, explicitly broken by the quantum anomaly, is equally vital for understanding non-perturbative QCD phenomena, such as the large mass of the $\eta'$ meson \cite{Weinberg:1975ui}. The restoration of these symmetries during the transition from hadronic matter to the quark-gluon plasma (QGP) provides critical insights into the phase structure of QCD and the properties of matter in the early universe.

A powerful framework for probing these symmetry restorations is provided by meson and topological susceptibilities. The symmetry breakings in QCD are directly reflected in meson susceptibility functions, which are defined as the two-point functions of quark bilinear operators in the low-energy limit \cite{Cohen:1996ng,Aoki:2012yj,HotQCD:2012vvd,Bhattacharya:2014ara,Aoki:2021qws, Bazavov:2019www, Ding:2025pbu, GomezNicola:2016ssy, GomezNicola:2017bhm, Cui:2021bqf, Cui:2022vsr,Giordano:2024jnc}. At high temperatures, as chiral symmetry is restored, the masses and susceptibilities of mesons related by chiral transformations become degenerate. This degeneracy allows meson susceptibilities to serve as a quantitative measure of chiral symmetry breaking and its restoration. Similarly, the effective restoration of the $U(1)_A$ symmetry can be quantified by the degeneracies between meson susceptibility functions connected by the $U(1)_A$ transformation.

The role of the $U(1)_A$ anomaly can be further quantified through the topological susceptibility, which is sensitive to fluctuations of the topological charge in the QCD vacuum. By making use of the Ward-Takahashi identity (WTI) associated with chiral symmetry, one can show that the topological susceptibility is correlated with the chiral and $U(1)_A$ partner structures in the meson susceptibility functions~\cite{Kawaguchi:2020kdl, Kawaguchi:2020qvg, Kawaguchi:2023olk,GomezNicola:2016ssy,GomezNicola:2017bhm, Cui:2021bqf}. This establishes the topological susceptibility as a key indicator for the breaking strength of the $U(1)_A$ symmetry through the chiral phase transition. Lattice QCD simulations at physical quark masses confirm that the magnitude of the topological susceptibility smoothly decreases at high temperatures~\cite{Borsanyi:2016ksw, Bonati:2015vqz, Petreczky:2016vrs,Athenodorou:2022aay}, providing direct evidence of the persistence anomaly and its subsequent suppression.

The restoration of chiral symmetry and the U(1)$_A$ symmetry are suggested to be connected, although the precise nature of this connection, especially its dependence on the number of quark flavors, is still not fully understood. A strong correlation between the topological susceptibility and chiral restoration has been extensively studied through meson susceptibility functions, both within 2-flavor QCD~\cite{Cohen:1996ng,Aoki:2012yj,Tomiya:2016jwr,Aoki:2020noz} and $2+1$-flavor QCD~\cite{HotQCD:2012vvd,Buchoff:2013nra,Bhattacharya:2014ara}. Furthermore, the $U(1)_A$ anomaly contribution remains manifest in meson susceptibility functions at high temperatures even when the light quarks are tuned to be massless, highlighting its fundamental and non-trivial impact on the phase structure of QCD~\cite{Ding:2021jtn}.

Calculating the low-energy two-point correlation functions requires  a non-perturbative technique that lies beyond the reach of conventional perturbative QCD. One powerful non-perturbative approach is the Anti-de Sitter/Conformal Field Theory (AdS/CFT) correspondence, or gauge/gravity duality ~\cite{Maldacena:1997re,Witten:1998qj}. Holographic QCD models built on this duality are broadly categorized into two approaches. The top-down approach starts from a consistent string theory in higher dimensions and derives a lower-dimensional theory with QCD-like properties~\cite{Karch:2002sh,Sakai:2004cn}. In contrast, the bottom-up approach begins with established QCD phenomenology and constructs a higher-dimensional gravitational dual that reproduces these features \cite{Erlich:2005qh,DaRold:2005mxj,Karch:2006pv,Ghoroku:2005vt,Gubser:2008ny,Gubser:2008yx,Gursoy:2007cb,Gursoy:2007er}. A significant development in the bottom-up approach is the soft-wall AdS/QCD model ~\cite{Karch:2006pv}, where a specific dilaton configuration is introduced, which successfully generates the linear Regge trajectory of meson spectra and describes the spontaneous breaking of chiral symmetry in the chiral limit. In the soft-wall model, the phase structure has been studied at finite temperature and/or finite chemical potentials for different flavor systems~\cite{Colangelo:2011sr,Cui:2013zha,Bartz:2016ufc,Chelabi:2015cwn,Chelabi:2015gpc,Fang:2015ytf,Fang:2016nfj,Li:2016smq,Bartz:2017jku,Ahmed:2024rbj}. Moreover, the thermal properties and dynamics of light-flavor mesons have been investigated via the two-point correlation function, with particular emphasis on the pseudo-Goldstone bosons emerging from the chiral phase transition \cite{Cao:2021tcr, Cao:2022csq}.
In this paper, we employ a holographic framework to calculate meson and topological susceptibilities, intending to investigate the restoration patterns of chiral and $U(1)_A$ symmetries at finite temperatures. Our work aims to provide a qualitative understanding of these critical QCD phenomena from a gravitational dual perspective.

The paper is organized as follows. Firstly, we will provide a brief introduction to the soft-wall AdS/QCD models for a three-flavor system and present the equations of motion in Section \ref{sectionII}. Then, in section \ref{sectionIII}, we will derive the meson susceptibilities from the two-point correlation functions for scalar and pseudoscalar sectors.
 Additionally, we will obtain the topological susceptibilities in section \ref{sectionIV}. Referring to the results in section \ref{sectionV}, we will show the results of the order parameter (chiral condensate), screening mass, mesons, and topological susceptibilities.
 Finally, we will conclude the results in section \ref{sectionVI}.

\section{Soft-Wall Model at Finite Temperature}
\label{sectionII}
We begin with a brief review of the soft-wall holographic model. The model is defined in a five-dimensional Anti-de Sitter (AdS$_5$) spacetime \cite{Karch:2006pv}. Temperature is incorporated via a black hole geometry, described by a static background metric. Specifically, we adopt the AdS-Schwarzschild (AdS-SW) black hole solution \cite{Chelabi:2015cwn,Chelabi:2015gpc}:

\begin{equation}
ds^{2}= e^{2 A_{s}(z)} \left( f(z) dt^{2} - \frac{1}{f(z)} dz^{2} - dx_{i} dx^{i}\right),
\label{metric}
\end{equation}
where $z$ is the holographic (fifth) coordinate. The scale function is given by $A_{s}(z)= - \log(z/L)$, scaled by the AdS curvature radius $L$ (set to $L=1$ for simplicity). The blackening factor is:
\begin{equation}
f(z) = 1- \frac{z^{4}}{z_{h}^{4}},
\label{fz}
\end{equation}
with $z_h$ denoting the horizon location, defined by $f(z_h)=0$. The Hawking temperature, identified as the temperature of the dual QCD system, is given by:
\begin{equation}
T=\frac{1}{4 \pi} \left|\frac{d f(z)}{dz}\right|_{z = z{h} } = \frac{1}{\pi z_{h}}.
\label{hawking}
\end{equation}

\subsection{The 5D Holographic Action}
Following the bottom-up approach to holographic QCD, we employ a five-dimensional gauge theory with $U(N_f)_L\times U(N_f)_R$ symmetry, where $N_f$ is the number of quark flavors \cite{Erlich:2005qh,DaRold:2005mxj,Karch:2006pv}. The model includes bulk left- ($L_M$) and right- ($R_M$) handed gauge fields. To incorporate the chiral condensation operator, a bulk scalar field $X$ is introduced; it transforms in the bifundamental representation under the chiral symmetry. To break the conformal invariance, a dilaton background field $\Phi(z)$ is included, acting as a soft infrared (IR) cutoff \cite{Karch:2006pv}.

The five-dimensional action for the soft-wall model is:
\begin{equation}
\begin{aligned}
S_{M} &=\int d^{5} x \sqrt{g} e^{-\Phi}\left\{ \operatorname{Tr}\left[\left(D^{M} X\right)^{\dagger}\left(D_{M} X\right)-V(X) -\frac{1}{4 g_{5}^{2}}\left(L^{M N} L_{M N}+R^{M N} R_{M N}\right)\right]- \gamma \operatorname{Re}(det[X])
\right\},
\end{aligned}
\label{action}
\end{equation}
where $g \equiv \det(g_{MN}) = e^{5A_s(z)} = z^{-5}$ is the determinant of the metric, and $D^{M}$ is the covariant derivative which is $D^{M} X=\partial^{M} X - i L^{M} X + i X R^{M}$. The field strength tensors are $L_{MN}=\partial_M L_N -\partial_N L_M -i[L_M,L_N]$ and $R_{MN}=\partial_M R_N -\partial_N R_M -i[R_M,R_N]$. The gauge coupling $g_{5}$ is fixed by matching the vector current correlator to its QCD asymptotic form: $g_{5}^{2} = 12 \pi^{2} /N_{c}$. The parameter $\gamma$ controls the strength of the determinant term, $\mathrm{Re}(\det[X])$, which is anomalous under $U(1)_A$ but invariant under $SU(N_f)_L\times SU(N_f)_R$.

The scalar potential $V(X)$ is taken to be:
\begin{equation}
V(X)= m_{5}^{2} \operatorname{Tr}\left( X^{\dagger} X\right) + \lambda \operatorname{Tr}\left( |X|^{4}\right),
\label{potential}
\end{equation}
where the five-dimensional mass parameter, from the AdS/CFT dictionary, is $m_{5}^{2} = (\Delta - p)(\Delta + p - 4) = -3$ for $\Delta=3$ and $p=0$. The positive dimensionless coupling $\lambda$ is introduced to ensure a non-vanishing quark condensate in the chiral limit ($m_q \to 0$); without this term, the condensate would accidentally vanish \cite{Gherghetta:2009ac}.

In the soft-wall model, there are different ways to get a consistent meson spectrum with the experimental data and a good description of the chiral phase transition. Here we consider the following two cases:

\begin{enumerate}
    \item Case I: The specific profile of the dilaton field, which respects both the IR and UV behavior with a standard 5D mass \cite{Chelabi:2015cwn,Chelabi:2015gpc},

\begin{equation}
    \Phi(z)=- \mu_{1}^{2}z^2  + (\mu_{1}^{2} + \mu_{g}^{2}) z^{2} \text{tanh}(\mu_{2}^{2} z^{2}) \textbf{   } \&  \textbf{   }  m_{5}^{2}=-3 .
\end{equation}
    
    \item Case II: An IR-modified soft-wall AdS/QCD model with a modified 5D mass and quadratic dilaton field \cite{Fang:2016nfj}\footnote{While other modifications that likewise introduce an additional scale to characterize chiral dynamics are conceivable (see, e.g., Refs. \cite{Rinaldi:2022dyh, Ballon-Bayona:2023zal}), the present discussion is restricted to the two classes outlined above.},
\begin{equation}
    \Phi(z)= \mu_{g}^{2} z^{2}  \textbf{   }  \& \textbf{   }  m_{5}^{2}=-3 - \mu_{c}^{2} z^{2}.
\end{equation}
 
\end{enumerate}

We analyze both variants of the model (Case I and Case II). To study chiral symmetry breaking, we consider the static, homogeneous expectation value of the scalar field $X$, which takes the form:
\begin{equation}
X(z)=X_0(z) = \mathrm{diag} \left( \frac{\chi_l(z)}{2}, \frac{\chi_l(z)}{2}, \frac{\chi_s(z)}{2}\right),
\label{chivev}
\end{equation}
where we impose isospin symmetry ($m_u = m_d \equiv m_l$). This implies the light quark components are equal: $\chi_u(z) = \chi_d(z) \equiv \chi_l(z)$, while $\chi_s(z)$ corresponds to the strange quark sector.

To isolate the dynamics of the chiral condensate, we set the bulk gauge fields to zero ($L_M = R_M = 0$). With this simplification, the full action \eqref{action} reduces to an effective action for the scalar VEVs, $\chi_l$ and $\chi_s$:
\begin{equation}
\begin{aligned}
S\left[\chi_l,\chi_s\right]=  \int d^5 x \sqrt{g} e^{-\Phi} \left\{g^{z z}\left(\frac{1}{2} \chi_l^{\prime 2} + \frac{1}{4} \chi_s^{\prime 2} \right) - \left(\frac{1}{2} m_{5}^{2}\chi_l^2+\frac{1}{4} m_{5}^{2}\chi_s^2+\frac{\lambda}{8} \chi_l^4 +\frac{\lambda}{16} \chi_s^4 + \frac{\gamma}{8}  \chi_l^2 \chi_s \right)\right\}
\end{aligned}
\label{actionchi}
\end{equation}
where $g^{zz} = -e^{-2A_s(z)} f(z)$ is the relevant component of the inverse metric.

\subsection{Equations of Motion}
Varying the effective action \eqref{actionchi} yields the following equations of motion (EOMs) for the fields $\chi_f(z)$ ($f = l, s$):
\begin{equation}
\begin{gathered}
    \chi_l^{\prime \prime}+\left(3 A_s^{\prime}-\Phi^{\prime}+\frac{f^{\prime}}{f}\right) \chi_l^{\prime}-\frac{e^{2 A_s}}{f}\left(m_{5}^{2}  + \frac{\gamma}{4} \chi_s  + \frac{\lambda}{2} \chi_l^2\right) \chi_l=0, \\
   \chi_s^{\prime \prime}+\left(3 A_s^{\prime}-\Phi^{\prime}+\frac{f^{\prime}}{f}\right) \chi_s^{\prime}-\frac{e^{2 A_s}}{f}\left(m_{5}^{2}  \chi_s  + \frac{\gamma}{4}  \chi_l^2  + \frac{\lambda}{2} \chi_s^3\right)=0,\\
\end{gathered}
\label{chieom3}
\end{equation}
where the prime ($^\prime$) denotes a derivative with respect to the holographic coordinate $z$.

The asymptotic behavior of the fields $\chi_l(z)$ and $\chi_s(z)$ in the ultraviolet (UV, $z \to 0$) and infrared (IR, $z \to z_h$) regions is derived for the two model variants, characterized by their distinct dilaton profiles $\Phi(z)$ and five-dimensional mass terms $m_5^2$. The expansions are as follows:

Case I:

\begin{equation}
\begin{gathered}
    \chi_{l}(z\to 0) = m_{l} \zeta z - \frac{\gamma}{4} m_{l} m_{s} \zeta^{2} z^2 + \frac{m_{l} \zeta}{4} \left(-4 \mu_{1}^{2} + m_{l}^{2} \zeta^{2} \lambda - \frac{\gamma^{2}}{8} (m_{l}^{2} \zeta^2 + m_{s}^{2} \zeta^2)  \right)  z^{3} \log(z) + \frac{\sigma_l}{\zeta} z^{3} ,\\
   \chi_{l}(z\to z_h)  = c_{l,0} - \frac{c_{l,0} ( -2 c_{l,0}^{2} \lambda +12 - \gamma c_{s,0})}{16 z_h} (z_h -z) + \mathcal{O}[(z_h -z)^{2}] ,\\
\end{gathered}
\label{UV3l1}
\end{equation}

\begin{equation}
\begin{gathered}
    \chi_{s}(z\to 0) = m_{s} \zeta z - \frac{\gamma}{4} m_{l}^{2} \zeta^{2} z^2 + \frac{1}{4} \left(-4 \mu_{1}^{2} m_{s} \zeta + m_{s}^{3} \zeta^{3} \lambda - \frac{\gamma^{2}}{4} m_{l}^{2}  m_{s} \zeta^3  \right)  z^{3} \log(z) + \frac{\sigma_s}{\zeta} z^{3} ,\\
   \chi_{s}(z\to z_h)  = c_{s,0} - \frac{( -2 c_{s,0}^{3} \lambda + 12 c_{s,0} - \gamma c_{l,0}^{2} )}{16 z_h} (z_h -z) + \mathcal{O}[(z_h -z)^{2}] ,\\
\end{gathered}
\label{UV3s1}
\end{equation}

Case II:

\begin{equation}
\begin{gathered}
    \chi_{l}(z\to 0) = m_{l} \zeta z - \frac{\gamma}{4} m_{l} m_{s} \zeta^{2} z^2 + \frac{m_{l} \zeta}{4} \left(4 \mu_{g}^{2} - 2 \mu_{c}^{2} + m_{l}^{2} \zeta^{2} \lambda - \frac{\gamma^{2}}{8} (m_{l}^{2} \zeta^2 + m_{s}^{2} \zeta^2)  \right)  z^{3} \log(z) + \frac{\sigma_l}{\zeta} z^{3} ,\\
   \chi_{l}(z\to z_h)  = c_{l,0} - \frac{c_{l,0} (4 \mu_{c}^{2} z_{h}^{2} -2 c_{l,0}^{2} \lambda +12 - \gamma c_{s,0})}{16 z_h} (z_h -z) + \mathcal{O}[(z_h -z)^{2}] ,\\
\end{gathered}
\label{UV3l2}
\end{equation}

\begin{equation}
\begin{gathered}
    \chi_{s}(z\to 0) = m_{s} \zeta z - \frac{\gamma}{4} m_{l}^{2} \zeta^{2} z^2 + \frac{1}{4} \left(4 \mu_{g}^{2} m_{s} \zeta - 2 \mu_{c}^{2} m_{s} \zeta + m_{s}^{3} \zeta^{3} \lambda - \frac{\gamma^{2}}{4} m_{l}^{2}  m_{s} \zeta^3  \right)  z^{3} \log(z) + \frac{\sigma_s}{\zeta} z^{3} ,\\
   \chi_{s}(z\to z_h)  = c_{s,0} - \frac{(4 c_{s,0} \mu_{c}^{2} z_{h}^{2} -2 c_{s,0}^{3} \lambda + 12 c_{s,0} - \gamma c_{l,0}^{2} )}{16 z_h} (z_h -z) + \mathcal{O}[(z_h -z)^{2}] ,\\
\end{gathered}
\label{UV3s2}
\end{equation}

where $m_l$ and $m_s$ are the light and strange quark masses, respectively, $\sigma_l$ and $\sigma_s$  are the chiral condensate of light and strange quark, $\zeta=\frac{\sqrt{N_c}}{2\pi}$ is a normalization factor which is fixed by matching the two point correlation function of the scalar operator with the 4D results \cite{Cherman:2008eh}, and $c_{l(s),0}$ is an integration constant.

\section{ Meson susceptibility in soft-wall model}
\label{sectionIII}
We consider two models in our work, for which the mass spectra of the mesons are given in Ref. \cite{Bartz:2016ufc} for case I and Refs. \cite{ Fang:2016nfj,Fang:2019lmd} for case II. Here, we focus on the chiral and $U(1)_A$ symmetry partners of the scalar and pseudoscalar mesons.

\subsection{Scalar sector}

For the scalar and pseudo-scalar mesons, the perturbations on the background have the following form,

\begin{equation}
X= (X_0 +S)  e^{2 i \pi},
\label{Xfield}
\end{equation}
with S representing scalar field fluctuation and $\pi$ pseudoscalar field fluctuation. The scalar field is defined by a $3 \times 3$ matrix,

\begin{equation}
S=S^a t^a=\frac{1}{\sqrt{2}}\left(\begin{array}{ccc}
\frac{a_0}{\sqrt{2}}+\frac{\sigma_8}{\sqrt{6}}+\frac{\sigma_0}{\sqrt{3}} & a^{+} & \kappa^{+} \\
a^{-} & -\frac{a_0}{\sqrt{2}}+\frac{\sigma_8}{\sqrt{6}}+\frac{\sigma_0}{\sqrt{3}} & \kappa^0 \\
\kappa^{-} & \bar{\kappa}^0 & -\frac{2 \sigma_8}{\sqrt{6}}+\frac{\sigma_0}{\sqrt{3}}
\end{array}\right),
\end{equation}
where $t^{a}=\lambda_{a}/2  (a=0,1,...,8)$ are the generators of $U(3)$, where $\lambda_{a=1,...,8}/2$ are the Gell-Mann matrices with $\lambda_{0}=\sqrt{2/3} \mathbf{1}_{3 \times 3} $. The $\sigma_0$ and $\sigma_8$ are referred to as the admixtures of the $f_{0}(500)$ and $f_{0}(980)$.

The effective action of the scalar fluctuation up to the second order is given by 

\begin{equation}
S_{\mathrm{S}}=\frac{1}{2} \int d^5x \sqrt{g} e^{-\Phi}\left[g^{\mu \nu} \partial_\mu S^{a} \partial_\nu S^{b} +g^{z z}\left(\partial_z S^{a}\right)\left(\partial_z S^{b}\right)-m_5^2 S^{a} S^{b}-M^{a,b}(z) S^{a} S^{b}\right] ,
\end{equation}
where

\begin{equation}
\begin{gathered}
    M^{a,b}(z)= \lambda M^{a,b}_{s}(z) +2 \gamma M^{a,b}_{det} (z),\\
    M^{a,b}_{s}(z)=2 Tr \left( \{t^{a},X_0\}\{t^{b},X_0\} +2 t^{a} t^{b} X_0 X_0  \right),
\end{gathered}
\label{massscalar}
\end{equation}
The non-zero values of the $M^{a,b}_{s}(z)$ and $M^{a,b}_{det} (z)$ are given in Appendix \ref{appendA}. For the sake of convenience, one can transform the system from the coordinate space (x) to the momentum space (p) by taking the Fourier transformation $S^{a}(\mathrm{x}, z)=\frac{1}{(2 \pi)^4} \int d^4 \mathrm{p} e^{i \mathrm{p.x}} S^{a}(\mathrm{p}, z)$. Here, we investigate the two-point correlation function and susceptibility of the $a_0(980)$ and $f_0(500)$ ($\sigma$) mesons in the scalar sector. First, let's consider the case of the $a_{0}(980)$ meson, where the flavor index $a=b=1,2$ or $3$. Then, one can obtain the EOM as follows

\begin{equation}
S^{a\prime \prime}+\left(3 A^{\prime}+\frac{f^{\prime}}{f}-\Phi^{\prime}\right) S^{a\prime}-\left(\frac{p^2}{f}+\frac{ m_5^2+  M^{a,a}(z)}{ f} A^{\prime 2}\right) S^{a}=0 ~.
\end{equation}

It is worth mentioning that this equation can not be solved analytically. Therefore, we need a numerical technique, such as the shooting method, to solve the EOM with the constraints from the boundary and horizon. Substituting the value of $M^{3,3}(z)$, the equation of motion reduces to \footnote{We omitted the flavor index for simplicity.}

\begin{equation}
S^{\prime \prime}+\left(3 A^{\prime}+\frac{f^{\prime}}{f}-\Phi^{\prime}\right) S^{\prime}-\left(\frac{p^2}{f}+\frac{4 m_5^2- \gamma \chi_s +6 \lambda \chi_l^2}{4 f} A^{\prime 2}\right) S=0 ~.
\end{equation}

The asymptotic solution of the scalar field $S(p, z)$ near the UV boundary $z\to 0$ is obtained as 

\begin{equation}
S(z \rightarrow 0)=s_1 z+\frac{\gamma}{4} \zeta m_s s_1 z^2+s_3 z^3-\frac{1}{32} \left[s_1\left(16\left(-p^2+\mu_c^2-2 \mu_g^2\right)-24 \zeta^2 \lambda m_l^2\right) +  s_1 \gamma^2 \zeta^{2} m_s^{2} \right] z^3 \log (z)+\mathcal{O}\left(z^4\right),
\end{equation}
where $s_1$ and $s_3$ are the integration constants. According to the holographic dictionary, $s_1$ corresponds to the external source $J_S$, and $s_3$ corresponds to the operator $\bar{q} q$.
Near the horizon $z=z_h$, one can also get the non-singular solution as

\begin{equation}
S\left(z \rightarrow z_h\right)=s_{h 0} - \frac{12-6 c_{l,0}^2 \lambda + \gamma c_{s,0} -4 p^2 z_h^2+4 \mu_c^2 z_h^2}{16 z_h} s_{h 0}\left(z_h-z\right)+\mathcal{O}\left[\left(z_h-z\right)^2\right],
\end{equation}
where $s_{h 0}$ is another integration constant. The parameters $c_{l,0}$ and $c_{s,0}$ are the integration constants of the $\chi_l$ and $\chi_s$ at the horizon, which can be achieved by solving Eq. \eqref{chieom3} numerically for different temperatures or $z_h$.
Substituting EOM into the action, one can get the corresponding on-shell action as

\begin{equation}
S_{\mathrm{a_0}}^{\text {on }}=-\left.\frac{1}{2} \int d^4 p f(z) S(-p, z) e^{3 A(z)-\Phi(z)} S^{\prime}(p, z)\right|_{z=\epsilon} ^{z=z_h},
\end{equation}
where $\epsilon$ is an UV cutoff regularizing the on-shell action. Finally, one can derive the two-point Green's function of the scalar meson by taking the second-order derivative of the on-shell action $S_{\mathrm{S}}^{\text {on }}$ with respect to the external source $J_S$

\begin{equation}
G_{\mathrm{a_0}}(p)=\left.\frac{\delta^2 S_{\mathrm{a_0}}^{\mathrm{on}}}{\delta J_{a_0}^* \delta J_{a_0}}\right|_{z=\epsilon}=-\frac{4 s_3}{s_1}-\frac{3 \gamma}{32} (\zeta m_s)^2  -\frac{3}{4} \zeta^2 \lambda m_l^2+\frac{1}{2}\left(\mu_c^2-2 \mu_g^2-p^2\right)
\label{Ga0}
\end{equation}

The scalar susceptibility is defined by the two-point Green's function at zero momentum transfer,

\begin{equation}
\chi_{a_0} = -  \lim_{p^{2}\to 0} G_{\mathrm{a_0}}(p) =-  \lim_{p^{2}\to 0}\left.\frac{\delta^2 S_{\mathrm{a_0}}^{\mathrm{on}}}{\delta J_{a_0}^* \delta J_{a_0}}\right|_{z=\epsilon}=\frac{4 s_3}{s_1}+\frac{3 \gamma}{32} (\zeta m_s)^2  +\frac{3}{4}  \lambda \zeta^2 m_l^2-\frac{1}{2}\left(\mu_c^2-2 \mu_g^2\right)
\end{equation}

We did the analysis for case II; however, to extend the following analysis to case I, remove $\mu_{c}$ and make the substitution $\mu_g^2 \rightarrow -\mu_{1}^2$ in Eqs. (22-26). This same transformation applies to all subsequent results.
 
 The mass-like parameter $M^{a,b}(z)$ in Eq. \eqref{massscalar} is non-diagonal, and there is a mixing between the singlet and octet states. In order to find the equation of motion for the physical states $\sigma$ and $f_0(980)$, we need to diagonalize the mass term $M^{a,b}(z)$ by applying an orthogonal transformation in the following way \cite{Kawaguchi:2020qvg}:
 
\begin{equation}
\begin{aligned}
\tilde{S}^{i} & =O^{i a} S^{a}, \\
\tilde{M}^{i,j} & =O^{i a}\left(M^{a, b}\right) O^{b i} .
\end{aligned}
\end{equation}

The value of $M^{a,b}(z)$ for the singlet, octet states with their mixing are

\begin{equation}
\begin{aligned}
 &   M^{0,0}= \lambda M^{0,0}_{s} + \gamma M^{0,0}_{det}=\frac{\lambda}{2}(2 \chi_{l}^{2} + \chi_{s}^{2}) + \frac{\gamma}{6} (2 \chi_{l} + \chi_{s}),\\
 &   M^{8,8}= \lambda M^{8,8}_{s} + \gamma M^{8,8}_{det}=\frac{\lambda}{2}( \chi_{l}^{2} +2 \chi_{s}^{2}) - \frac{\gamma}{12} (4 \chi_{l} - \chi_{s}),\\
 &   M^{0,8}= \lambda M^{0,8}_{s} + \gamma M^{0,8}_{det}=\frac{\lambda}{2}( \chi_{l}^{2} - \chi_{s}^{2}) - \frac{\gamma}{3\sqrt{2}} (4 \chi_{l} - \chi_{s}).\\
&
\end{aligned}
\end{equation}

The corresponding mass like parameter for the physical states $\sigma$ and $f_{0}(980)$ are obtained as

\begin{equation}
\begin{aligned}
& M_{\sigma}(z)=M^{0,0}(z) \cos ^2 \theta_S+M^{8,8}(z) \sin ^2 \theta_S + 2 M^{0,8}(z) \cos \theta_S \sin \theta_S, \\
& M_{f_0(980)}(z)=M^{0,0}(z) \sin ^2 \theta_S+M^{8,8}(z) \cos ^2 \theta_S - 2 M^{0,8}(z) \cos \theta_S \sin \theta_S, \\
&
\end{aligned}
\end{equation}
where $\theta_S$ represents the scalar mixing angle,
\begin{equation}
\tan 2 \theta_S(z)=\frac{2 M^{0,8}(z)}{M^{0,0}(z)-M^{8,8}(z)} .
\end{equation}
 The difference between the mixing angle here with the conventional linear sigma model is that, the $\theta_S$ is a function of the holographic coordinate $z$. Now, we can obtain the equation of motion for $\sigma$ from the action,

\begin{equation}
S_{\mathrm{\sigma}}=\frac{1}{2} \int d x^5 \sqrt{g} e^{-\Phi}\left[g^{\mu \nu} \partial_\mu S \partial_\nu S+g^{z z}\left(\partial_z S\right)^2-m_5^2 S^2-M_{\sigma}(z) S^2\right] .
\end{equation}

\begin{equation}\label{pscalareq}
S^{\prime \prime}+\left(3 A^{\prime}+\frac{f^{\prime}}{f}-\Phi^{\prime}\right) S^{\prime}-\left(\frac{p^2}{f}+\frac{ m_5^2+ M_{\sigma}(z)}{ f} A^{\prime 2}\right) S=0
\end{equation}

Similar to the $a_0$ state, we can obtain the asymptotic solution for $\sigma$ near the boundary and the horizon. The solution near boundary is obtained as 

\begin{equation}
S(z \rightarrow 0)=s_1 z + s_1 s_{\gamma} z^2     - \frac{1}{2}s_1 \left(\mu_c^2-2 \mu_g^2 -p^2 + s_{\gamma}^2 + s_{\lambda}\right) z^3 \log (z)+s_3 z^3+\mathcal{O}\left(z^4\right),
\end{equation}
where

\begin{equation}
    \begin{aligned}
    & s_{\gamma}= \gamma \zeta  \left( \frac{1}{12} (4 m_l-m_s) (\sin \theta_S(\epsilon))^2 -\frac{1}{6} (2 m_l+m_s) (\cos \theta_S(\epsilon))^2  + \frac{1}{3\sqrt{2}} (m_l-m_s) \sin \theta_S(\epsilon)  \cos \theta_S(\epsilon)      \right), \\
    & s_{\lambda}= \lambda \zeta^2 \left(  \frac{1}{2} ( m_l^2 +2 m_s^2) (\sin \theta_S(\epsilon))^2 + \frac{1}{2} (2 m_l^2 + m_s^2) (\cos \theta_S(\epsilon))^2 + \sqrt{2} (m_l^2 - m_s^2) \sin \theta_S(\epsilon)  \cos \theta_S(\epsilon)     \right).\\
    &
    \end{aligned}
\end{equation}

And near to the horizon, the solution is 

\begin{equation}
S\left(z \rightarrow z_h\right)=s_{h 0} - \frac{3 - M_{\sigma}(zh) - p^2 zh^2}{4 z_h} s_{h 0}\left(z_h-z\right)+\mathcal{O}\left[\left(z_h-z\right)^2\right].
\end{equation}

The two-point correlation function and the $\sigma$ susceptibility obtained from the second-order functional derivative of the on-shell action for $\sigma$ with respect to the source are as follows

\begin{equation}
G_{\mathrm{\sigma}}(p)=\left.\frac{\delta^2 S_{\mathrm{\sigma}}^{\mathrm{on}}}{\delta J_{\sigma}^* \delta J_{\sigma}}\right|_{z=\epsilon}=-\frac{4 s_3}{s_1} - 2 s_{\gamma}^2  + \frac{1}{2} \left(\mu_c^2-2 \mu_g^2 -p^2 + s_{\gamma}^2 + s_{\lambda}\right).
\end{equation}

\begin{equation}
\chi_{\sigma} = -  \lim_{p^{2}\to 0} G_{\mathrm{\sigma}}(p) =\frac{4 s_3}{s_1} + 2 s_{\gamma}^2  - \frac{1}{2} \left(\mu_c^2-2 \mu_g^2+ s_{\gamma}^2 + s_{\lambda}\right).
\label{chisig}
\end{equation}


\subsection{Pseudoscalar channel}

In this section, we investigate the two-point correlation function and susceptibilities of the pion and $\eta$ particles. The pseudoscalar field is defined through the perturbation on the background as given by Eq. \eqref{Xfield}, where $\pi$ is the pseudoscalar field and is given by 

\begin{equation}
\pi= \pi^a t^a=\frac{1}{\sqrt{2}}\left(\begin{array}{ccc}
\frac{\pi^0}{\sqrt{2}}+\frac{\eta_8}{\sqrt{6}}+\frac{\eta_0}{\sqrt{3}} & \pi^{+} & K^{+} \\ 
\pi^{-} & -\frac{\pi^0}{\sqrt{2}}+\frac{\eta_8}{\sqrt{6}}+\frac{\eta_0}{\sqrt{3}} & K^0 \\
K^{-} & \bar{K}^0 & -\frac{2 \eta_8}{\sqrt{6}}+\frac{\eta_0}{\sqrt{3}}
\end{array}\right),
\end{equation}
where $\eta^0$ and $\eta_8$ are singlet and octet components of the pseudoscalar field and referred to as the mixing of the physical $\eta$ and $\eta^\prime$ particles. It is worth mentioning that, the axial-vector field contributes to the pseudoscalar sectors and couples to the pion field. 

The pion field and the longitudinal part ($\varphi$) of the
axial-vector field are coupled in the pseudoscalar channel. The pion fluctuation becomes 

\begin{equation}
\begin{aligned}
S_\pi= & -\frac{1}{2 g_5{ }^2} \int d^5 x \sqrt{g} e^{-\Phi} \sum_{i=1}^3\left\{g^{\mu \nu} g^{z z} \partial_z \partial_\mu \varphi^i \partial_z \partial_\nu \varphi^i-g_5{ }^2 \chi_l^2\left(g^{\mu \nu} \partial_\mu \varphi^i \partial_\nu \varphi^i\right.\right. \\
& \left.\left.+g^{\mu \nu} \partial_\mu \pi^i \partial_\nu \pi^i+g^{z z}\left(\partial_z \pi^i\right)^2-2 g^{\mu \nu} \partial_\mu \varphi^i \partial_\nu \pi^i\right)\right\} .
\end{aligned}
\end{equation}

From the action, one can derive the EOMs for the pion field and the axial-vector field as
\begin{equation}
\begin{aligned}
\varphi^{\prime \prime}+\left(A^{\prime}+\frac{f^{\prime}}{f}-\Phi^{\prime}\right) \varphi^{\prime}-\frac{e^{2 A} g_5^2 \chi_l^2}{f}(\varphi-\pi) & =0, \\
\pi^{\prime \prime}+\left(3 A^{\prime}+\frac{f^{\prime}}{f}-\Phi^{\prime}+\frac{2 \chi_l^{\prime}}{\chi_l}\right) \pi^{\prime}+\frac{p^2}{f}(\varphi-\pi) & =0 .
\end{aligned}
\label{eom}
\end{equation}

The asymptotic solutions of the EOMs for the pion field at the boundary can be easily derived as \cite{Cao:2021tcr},

\begin{equation}
\begin{aligned}
& \varphi(z \rightarrow 0)=c_f-\frac{1}{2} \zeta^2 g_5^2 m_l^2 \pi_0 z^2 \log (z)+\varphi_2 z^2+\mathcal{O}\left(z^3\right), \\
& \pi(z \rightarrow 0)=\pi_0+c_f + \frac{1}{2} \pi_0 p^2 z^2 \log (z)+\pi_2 z^2+\mathcal{O}\left(z^3\right),
\end{aligned}
\end{equation}
where $c_f, \varphi_2, \pi_0$, and $\pi_2$ are the integration constants. $\pi_0$ is identified as the external source $J_\pi$. It has been pointed out by the work of Ref. \cite{Cao:2020ryx} that $c_f$ is a redundant free parameter and can be set to zero for simplicity. On the other hand, we can also derive the boundary conditions at the horizon, which take the following forms,

\begin{equation}
\begin{aligned}
& \varphi\left(z \rightarrow z_h\right)=-\frac{c_{l,0}^2 \pi^2}{z_h} \pi_{h 0}\left(z_h-z\right)+\mathcal{O}\left[\left(z-z_h\right)^2\right], \\
& \pi\left(z \rightarrow z_h\right)=\pi_{h 0}+\frac{p^2 z_h}{4} \pi_{h 0}\left(z_h-z\right)-\mathcal{O}\left[\left(z-z_h\right)^2\right] .
\end{aligned}
\end{equation}

Here, $\pi_{h 0}$ is another integration constant. The on-shell action of the pion part is

\begin{equation}
S_\pi^{\mathrm{on}}=-\left.\frac{1}{2 g_5^2} \int d^4 p e^{A-\Phi}\left[e^{2 A} g_5^2 f \chi_l^2 \pi(-p, z) \pi^{\prime}(p, z)+p^2 f \varphi(-p, z) \varphi^{\prime}(p, z)\right]\right|_{z=\epsilon} ^{z=z_h}.
\end{equation}



From Eq. \eqref{eom}, we can find a first order differential equation for $\pi$ and $\varphi$ fields as

\begin{equation}
       p^{2} \partial_{z} \varphi + e^{2 A(z)} g_{5}^{2} \chi_l^{2} \partial_{z} \pi=0
\end{equation}

Keeping in mind that, one can write the fields in holographic QCD as the source and the bulk-to-boundary propagator, we can write both $\pi$ and $\varphi$ near the boundary as

\begin{equation}
\begin{aligned}
& \varphi(z \rightarrow 0)=\pi_0 \varphi_b=\pi_0 \left(-\frac{1}{2} \zeta^2 g_5^2 m_l^2 z^2 \log (z)+\frac{\varphi_2}{\pi_0} z^2+\mathcal{O}\left(z^3\right) \right), \\
& \pi(z \rightarrow 0)=\pi_0 \pi_b =\pi_0 \left(1+ \frac{1}{2} p^2 z^2 \log (z)+ \frac{\pi_2}{\pi_0} z^2+\mathcal{O}\left(z^3\right) \right),
\end{aligned}
\end{equation}
where $b$ refers to the bulk-to-boundary propagator. We may construct a solution of the pion field perturbatively in $m_{\pi}$ by letting $\varphi_b(z)=a_1(0,z)-1$ \cite{Erlich:2005qh}, where $a_1$ is the bulk-to-boundary propagator of the axial-vector field for the $a_1$ meson with $p^{2}=0$. Now, the on-shell action for the pion becomes

\begin{equation}
S_\pi^{\mathrm{on}}=-\left.\frac{1}{2 g_5^2} \int d^4 p e^{A-\Phi} \pi_0 \pi_0 \left[- f m_{\pi}^{2} \pi_b(-p, z)  a_1^{\prime}(0, z) +p^2 f \varphi_b(-p, z) \varphi_b^{\prime}(p, z)\right]\right|_{z=\epsilon} ^{z=z_h},
\label{actionpi}
\end{equation}
where $m_{\pi}^2=-\bold{p}^2$ is the screening mass of the pion. Now, before going to find the two-point correlation function, we need to find the solution of the $a_1$ field. Considering the gauge field in the action Eq. \eqref{action}, we can obtain the axial-vector fluctuation of the $a_1$ particle as
\begin{equation}
\begin{gathered}
S_{a_1}=-\frac{1}{2 g_5^2} \int d^5 x \sqrt{g} e^{-\Phi}\left\{\sum_{i=1}^3\left\{g^{z z} g^{\mu \nu} \partial_z a_{1, \mu}^i \partial_z a_{1, \nu}^i+g^{\mu \nu} g^{m n} \partial_\mu a_{1, m}^i \partial_\nu a_{1, n}^i\right\}\right. \\
\left.-g_5^2 \chi^2 \sum_{i=1}^3 g^{m n} a_{1, m}^i a_{1, n}^i\right\} .
\end{gathered}
\end{equation}
 With the equation of motion derived from the action 
\begin{equation}
a_1^{\prime \prime}+\left(A^{\prime}+\frac{f^{\prime}}{f}-\Phi^{\prime}\right) a_1^{\prime}-\frac{e^{2 A} g_5^2 \chi^2+p^2}{f} a_1=0.
\label{axial}
\end{equation}

Similarly, we obtain the UV boundary conditions for the bulk-to-boundary propagator of the axial-vector,
\begin{equation}
a_1(0,z \rightarrow 0)=1+\frac{a_{1,2}}{a_{1,0}} z^2+\frac{1}{2}  z^2 \log (z)\left[\zeta^2 g_5^2 m_l^2\right]+\mathcal{O}\left(z^3\right),
\end{equation}
where $a_{1,0}$ (corresponds to the source of the axial-vector field) and $a_{1,2}$ are the integration constants. Inserting the asymptotic value of $a_1$ field into Eq. \eqref{actionpi}, one can obtain the two-point correlation function of the pion

\begin{equation}
G_{\mathrm{\pi}}(p) \propto\frac{m_{\pi}^{2} }{2 g_{5}^{2}} \left[\frac{2  a_{1,2}}{a_{1,0}} + \frac{1}{2} (\zeta^2 g_5^2 m_l^2 )   \right].
\end{equation}

Then, following the standard definition of the susceptibility, we can obtain the pion susceptibility as 

\begin{equation}
\chi_{\pi} = - \lim_{p^{2}\to 0} G_{\mathrm{\pi}}(p) \propto- \frac{m_{\pi}^{2} }{2 g_{5}^{2}} \left[\frac{2  a_{1,2}}{a_{1,0}} + \frac{1}{2} \zeta^2 g_5^2 m_l^2   \right].
\label{susc}
\end{equation}

By using the definition of the pion decay constant in holographic QCD, $ f_\pi^2=-\frac {e^A}{g_5^2} a_1^{\prime}(0, z)|_{z=\epsilon}$, we can see that the pion susceptibility is 

\begin{equation}
\chi_{\pi} \propto  \frac{1 }{2  } m_{\pi}^{2} f_\pi^2.
\label{suscfpi}
\end{equation}

Generally, following Ref. [53], when the source $J_{\pi}$ and the operator $O_{\pi^a} = \bar{q} \gamma_{5} t^a q$ are defined through the boundary coefficients of the bulk fields $\pi$ and $\varphi$, there exists a freedom to redefine them as
\[
    J_{\pi} = \kappa \, \pi_{0}, \qquad O_{\pi} = \frac{\pi_{2}}{\kappa},
\]
with a rescaling factor $\kappa$, which preserves the coupling form $J_{\pi} O_{\pi}$ in the partition function. This redefinition is typically neglected when analyzing the pole structure of Green's functions. However, in a quantitative study of the susceptibility, such a redefinition introduces an overall factor of $\kappa^{2}$ in the definition of the Green's function, which must be fixed by four-dimensional general relations. This is precisely why proportional signs are used in the above equations.

To determine the precise prefactor of the susceptibility, we invoke the Ward--Takahashi identity (WTI). As shown in Ref.~\cite{Kawaguchi:2023olk}, the pion susceptibility $\chi_{\pi}$ derived from the QCD Lagrangian is given by

\begin{equation}
    \chi_{\pi}=i \frac{\left< \bar{q} q \right>}{m_{l}}.
\label{WI}
\end{equation}

By using the GOR relation $2 m_l \left< \bar{q} q\right>_l = m_{\pi}^{2} f_{\pi}^{2}$, we can match Eq. \eqref{suscfpi} with the one found from the WTI in Eq. \eqref{WI}, and conclude that we need to normalize Eq. \eqref{suscfpi} with $1/(m_{l}^{2})$,

\begin{equation}
\chi_{\pi} = \frac{1 }{2 m_l^2  } m_{\pi}^{2} f_\pi^2.
\end{equation}

Similar to the scalar channel, there is a mixing between the singlet and octet channels of the pseudoscalar field. The action for the singlet and octet channels is given by

\begin{equation}
\begin{aligned}
S_\pi= & -\frac{1}{2 g_5{ }^2} \int d^5 x \sqrt{g} e^{-\Phi} \left\{\sum_{i,j=0,8} \left\{g^{\mu \nu} g^{z z} \partial_z \partial_\mu \varphi^i \partial_z \partial_\nu \varphi^j-g_5{ }^2 M_{A}^{i,j}(z)\left(g^{\mu \nu} \partial_\mu \varphi^i \partial_\nu \varphi^j\right.\right.\right. \\
& \left.\left.\left.+g^{\mu \nu} \partial_\mu \pi^i \partial_\nu \pi^j+g^{z z} \partial_z \pi^i \partial_z \pi^j -2 g^{\mu \nu} \partial_\mu \varphi^i \partial_\nu \pi^j\right) \right\}  - g_5{ }^2\frac{3 \gamma}{4} \chi_l^2 \chi_s \pi^0 \pi^0 \right\} .
\end{aligned}
\label{actionpseudo}
\end{equation}

where $i$ and $j$ take the values of 0 and 8, and $M_{A}^{i,j}(z)$ is defined by

\begin{equation}
    M_{A}^{i,j}(z)=2 Tr\left(\{t^{i},X_0\}.\{t^{j},X_0\} \right).
\end{equation}

The mixing between $M_{A}^{0,0}(z)$ and $M_{A}^{8,8}(z)$ can be encountered by applying the orthogonal rotation such as 

\begin{equation}
\begin{aligned}
\tilde{\pi}^{i} & =O^{i a} \pi^{a} \\
\tilde{M}_{A}^{i,j} & =O^{i a}\left(M_{A}^{a, b}\right) O^{b i} .
\end{aligned}
\end{equation}

\begin{equation}
\begin{aligned}
 &   M_{A}^{0,0}(z)=\frac{1}{3} (2 \chi_l^{2}(z)+ \chi_s^{2}(z))\\
 &   M_{A}^{8,8}=\frac{1}{3} ( \chi_l^{2}(z)+2 \chi_s^{2}(z))\\
 &   M_{A}^{0,8}=\frac{\sqrt{2}}{3} ( \chi_l^{2}(z) - \chi_s^{2}(z))\\
&
\end{aligned}
\end{equation}

The mass term for the physical $\eta$ and $\eta^{\prime}$ states are become

\begin{equation}
\begin{aligned}
& M_{\eta^{\prime}}(z)=M_{A}^{0,0}(z) \cos ^2 \theta_p + M_{A}^{8,8}(z) \sin ^2 \theta_p + 2 M_{A}^{0,8}(z) \cos \theta_p \sin \theta_p, \\
& M_{\eta}(z)=M_{A}^{0,0}(z) \sin ^2 \theta_p + M_{A}^{8,8}(z) \cos ^2 \theta_p - 2 M_{A}^{0,8}(z) \cos \theta_p \sin \theta_p, \\
&
\end{aligned}
\end{equation}

where $\theta_p(z)$ is the pseudoscalar mixing angle given by

\begin{equation}
\tan 2 \theta_p(z)=\frac{2 M_{A}^{0,8}(z)}{M_{A}^{0,0}(z)-M_{A}^{8,8}(z)} .
\end{equation}

It is worth noting that in Eq. \eqref{actionpseudo}, only the singlet state contribute to the determinant term, and after applying the orthogonal transformation, there is a mixing term between $\eta$ and $\eta^{\prime}$ states which results in a coupled equation of motions for them. The equation of motion for $\eta$ and $\eta^{\prime}$
\begin{equation}
\begin{aligned}
&\varphi_{\eta}^{\prime \prime}+\left(A^{\prime}+\frac{f^{\prime}}{f}-\Phi^{\prime}\right) \varphi_{\eta}^{\prime}-\frac{e^{2 A} g_5^2  M_{\eta}(z)}{f}(\varphi_{\eta}-\pi_{\eta})  =0, \\
&\pi_{\eta}^{\prime \prime}+\left(3 A^{\prime}+\frac{f^{\prime}}{f}-\Phi^{\prime}+ \frac{(M_{\eta})^{\prime}(z)}{M_{\eta}(z)}\right) \pi_{\eta}^{\prime}+\frac{p^2  }{f}(\varphi_{\eta}-\pi_{\eta})+\frac{3 e^{2 A} \gamma}{4 M_{\eta}(z) f} \chi_l^2 \chi_s \left( \pi_{\eta} \sin ^2 \theta_p(z) - \pi_{\eta^{\prime}} \sin \theta_p(z) \cos \theta_p(z)  \right) =0 .\\
&
\end{aligned}
\label{eometa}
\end{equation}

\begin{equation}
\begin{aligned}
&\varphi_{\eta^{\prime}}^{\prime \prime}+\left(A^{\prime}+\frac{f^{\prime}}{f}-\Phi^{\prime}\right) \varphi_{\eta^{\prime}}^{\prime}-\frac{e^{2 A} g_5^2  M_{\eta^{\prime}}(z)}{f}(\varphi_{\eta^{\prime}}-\pi_{\eta^{\prime}})  =0, \\
&\pi_{\eta^{\prime}}^{\prime \prime}+\left(3 A^{\prime}+\frac{f^{\prime}}{f}-\Phi^{\prime}+ \frac{(M_{\eta^{\prime}})^{\prime}(z)}{M_{\eta^{\prime}}(z)}\right) \pi_{\eta^{\prime}}^{\prime}+\frac{p^2  }{f}(\varphi_{\eta^{\prime}}-\pi_{\eta^{\prime}})+\frac{3 e^{2 A} \gamma}{4 M_{\eta^{\prime}}(z) f} \chi_l^2 \chi_s \left( \pi_{\eta^{\prime}} \cos ^2 \theta_p(z) - \pi_{\eta} \sin \theta_p(z) \cos \theta_p(z)  \right) =0 .\\
&
\end{aligned}
\label{eometaprime}
\end{equation}

The expansion of the pseudoscalar field and the longitudinal part of the axial-vector field near the boundary is obtained as follows

\begin{equation}
\begin{aligned}
& \varphi_{\eta}(z \rightarrow 0)=c_f-\frac{1}{2} g_5^2 M_{\eta}(\epsilon) \pi_{\eta,0} z^2 \log (z)+\varphi_{\eta,2} z^2+\mathcal{O}\left(z^3\right), \\
& \pi_{\eta}(z \rightarrow 0)=\pi_{\eta,0}+c_f + \pi_{\eta,1} z +  \\
& \frac{1}{2} \left( \pi_0 p^2 - \left(\frac{3 \gamma \zeta^3 m_l^{2} m_s }{4 M_{\eta}(\epsilon)}\right) \left(\pi_{\eta,1} \sin ^2 \theta_p(\epsilon) -\pi_{\eta^{\prime},1} \cos \theta_p(\epsilon) \sin \theta_p(\epsilon) \right)  \right) z^2 \log (z)+\pi_{\eta,2} z^2+\mathcal{O}\left(z^3\right),
\end{aligned}
\end{equation}

\begin{equation}
\begin{aligned}
& \varphi_{\eta^{\prime}}(z \rightarrow 0)=c_f-\frac{1}{2} g_5^2 M_{\eta}(\epsilon) \pi_{\eta^{\prime},0} z^2 \log (z)+\varphi_{\eta^{\prime},2} z^2+\mathcal{O}\left(z^3\right), \\
& \pi_{\eta^{\prime}}(z \rightarrow 0)=\pi_{\eta^{\prime},0}+c_f + \pi_{\eta^{\prime},1} z +  \\
& \frac{1}{2} \left( \pi_0 p^2 - \left(\frac{3 \gamma \zeta^3 m_l^{2} m_s }{4 M_{\eta^{\prime}} (\epsilon)}\right)\left (\pi_{\eta^{\prime},1} \cos ^2 \theta_p(\epsilon) -\pi_{\eta,1} \cos \theta_p(\epsilon) \sin \theta_p(\epsilon) \right)  \right) z^2 \log (z)+\pi_{\eta^{\prime},2} z^2+\mathcal{O}\left(z^3\right),
\end{aligned}
\end{equation}
where

\begin{equation}
    \pi_{\eta,1}=\frac{-1}{p^2 +2 \mu_{g}^{2}} \frac{3 \gamma \zeta^3 m_l^{2} m_s }{4 M_{\eta}(\epsilon)} \left(\pi_{\eta,2} \sin ^2 \theta_p(\epsilon) - \pi_{\eta^{\prime},2} \cos \theta_p(\epsilon) \sin \theta_p(\epsilon) \right)
\end{equation}

\begin{equation}
    \pi_{\eta^{\prime},1}=\frac{-1}{p^2+2 \mu_{g}^{2}} \frac{3 \gamma \zeta^3 m_l^{2} m_s }{4 M_{\eta^{\prime}}(\epsilon)} \left(\pi_{\eta^{\prime},2} \cos ^2 \theta_p(\epsilon) -\pi_{\eta,2} \cos \theta_p(\epsilon) \sin \theta_p(\epsilon) \right)
\end{equation}

Meanwhile, the non-singular solutions near the horizon are  

\begin{equation}
\begin{aligned}
& \varphi_{\eta}\left(z \rightarrow z_h\right)=-\frac{\pi^2 M_{\eta}(zh)}{ z_h} \pi_{\eta,h 0}\left(z_h-z\right)+\mathcal{O}\left[\left(z-z_h\right)^2\right], \\
& \pi_{\eta}\left(z \rightarrow z_h\right)=\pi_{\eta,h 0}-\frac{\pi_{\eta,h 0} zh}{4} \left(p^2-\frac{3 \gamma  c_{l,0}^2 c_{s,0} }{4 M_{\eta}(zh) zh^2} \left(  \sin ^2 \theta_p(zh) - \frac{\pi_{\eta^{\prime},h 0}}{\pi_{\eta,h 0}} \sin \theta_p(zh) \cos \theta_p(zh)  \right)\right) \left(z_h-z\right)+\mathcal{O}\left[\left(z-z_h\right)^2\right] .
\end{aligned}
\end{equation}

\begin{equation}
\begin{aligned}
& \varphi_{\eta^{\prime}}\left(z \rightarrow z_h\right)=-\frac{\pi^2 M_{\eta^{\prime}}(zh)}{ z_h} \pi_{\eta^{\prime},h 0}\left(z_h-z\right)+\mathcal{O}\left[\left(z-z_h\right)^2\right], \\
& \pi_{\eta^{\prime}}\left(z \rightarrow z_h\right)=\pi_{\eta^{\prime},h 0}-\frac{\pi_{\eta^{\prime},h 0} zh}{4} \left(p^2-\frac{3 \gamma  c_{l,0}^2 c_{s,0} }{4 M_{\eta^{\prime}}(zh) zh^2} \left(  \cos ^2 \theta_p(zh) - \frac{\pi_{\eta,h 0}}{\pi_{\eta^{\prime},h 0}} \sin \theta_p(zh) \cos \theta_p(zh)  \right)\right) \left(z_h-z\right)+\mathcal{O}\left[\left(z-z_h\right)^2\right] .
\end{aligned}
\end{equation}

After finding the on-shell action and taking the second-order functional derivative with respect to the source, one can obtain the two-point Green's function. Finally, the $\eta$ susceptibility is given by the two-point Green's function at zero momentum transfer as mentioned earlier

\begin{equation}
\begin{aligned}
\chi_{\eta} = -  \lim_{p^{2}\to 0} G_{\mathrm{\eta}}(p)= & -\frac{M_{\eta}(\epsilon)}{2 g_5^2\pi_{\eta,0} \pi_{\eta,0}} \left[  \pi_{\eta,1}^2   - \pi_{\eta,0} \left(\frac{3 \gamma \zeta^3 m_l^{2} m_s }{4 M_{\eta}(\epsilon)}\right) \left(\pi_{\eta,1} \sin ^2 \theta_p(\epsilon) -\pi_{\eta^{\prime},1} \cos \theta_p(\epsilon) \sin \theta_p(\epsilon) \right) +  2 \pi_{\eta,0}\pi_{\eta,2} \right].
\end{aligned}
\end{equation}

Remark: The $\chi_{\eta}$ includes the contribution of both light and strange quarks. However, only the light quark component contributes to the study of $SU(2)$ chiral symmetry restoration and $U(1)_A$ symmetry restoration. Therefore, we can decompose the singlet and octet states of the pseudoscalar field in terms of the quark states as follows: 

\begin{equation}
    \begin{gathered}
     \pi^0 = \frac{1}{\sqrt{3}} (u \bar{u} + d \bar{d} + s \bar{s}) = \frac{\sqrt{2}}{\sqrt{3}} \pi^l +\frac{1}{\sqrt{3}} \pi^s , \\
     \pi^8 = \frac{1}{\sqrt{6}} (u \bar{u} + d \bar{d} - 2 s \bar{s}) = \frac{1}{\sqrt{3}} \pi^l - \frac{\sqrt{2}}{\sqrt{3}} \pi^s ,
    \end{gathered}
\label{qs}
\end{equation}
where 

\begin{equation}
    \begin{gathered}
     \pi^l = \frac{1}{\sqrt{2}} (u \bar{u} + d \bar{d} ) , \\
     \pi^s =  s \bar{s} .
    \end{gathered}
\end{equation}

We can express this decomposition in the language of orthogonal transformation as 

\begin{equation}
\binom{\pi^8}{\pi^0}=\left(\begin{array}{cc}
\cos \phi & -\sin \phi \\
\sin \phi & \cos \phi
\end{array}\right)\binom{\pi^l}{\pi^s},
\end{equation}
where $\phi=54.7^{\text{o}}$. Now one can write the mass term for the light and strange quark components as

\begin{equation}
\begin{aligned}
& M_{\pi^{l}}(z)=M_{A}^{8,8}(z) \cos ^2 \phi + M_{A}^{0,0}(z) \sin ^2 \phi + 2 M_{A}^{0,8}(z) \cos \phi \sin \phi, \\
& M_{\pi^s}(z)=M_{A}^{8,8}(z) \sin ^2 \phi + M_{A}^{0,0}(z) \cos ^2 \phi - 2 M_{A}^{0,8}(z) \cos \phi \sin \phi. \\
&
\end{aligned}
\end{equation}

The steps of deriving the $\eta_l$ susceptibility are similar to $\chi_{\eta}$. The $\chi_{\eta_l}$ is given by 

\begin{equation}
\begin{aligned}
\chi_{\eta_l} = -  \lim_{p^{2}\to 0} G_{\mathrm{\eta_l}}(p)= & -\frac{M_{\pi^l}(\epsilon)}{2 g_5^2 \pi_{\eta_l,0} \pi_{\eta_l,0}} \left[  \pi_{\eta_l,1}^2   - \pi_{\eta_l,0} \left(\frac{3 \gamma \zeta^3 m_l^{2} m_s }{4 M_{\pi^l}(\epsilon)}\right) \left(\pi_{\eta_l,1} \sin ^2 \phi +\pi_{\eta^{s},1} \cos \phi \sin \phi \right) +  2 \pi_{\eta_l,0}\pi_{\eta_l,2} \right],
\end{aligned}
\end{equation}
with

\begin{equation}
    \pi_{\eta_l,1}=\frac{-1}{2 \mu_{g}^{2}} \frac{3 \gamma \zeta^3 m_l^{2} m_s }{4 M_{\pi^l}(\epsilon)} \left(\pi_{\eta_l,2} \sin ^2 \phi + \pi_{\eta_s,2} \cos \phi \sin \phi \right)
\end{equation}

\begin{equation}
    \pi_{\eta_s,1}=\frac{-1}{2 \mu_{g}^{2}} \frac{3 \gamma \zeta^3 m_l^{2} m_s }{4 M_{\pi^s}(\epsilon)} \left(\pi_{\eta_s,2} \cos ^2 \phi +\pi_{\eta_l,2} \cos \phi \sin \phi \right).
\end{equation}

Keeping in mind that the normalization factor we have found for the pion susceptibility must be added to all pseudoscalar channels, including the $\eta_l$ susceptibility. 


\section{Topological susceptibility}
\label{sectionIV}

The topological susceptibility is a quantity measuring the topological charge fluctuation of the QCD-$\theta$ vacuum, which is defined as the curvature of the $\theta$-dependent QCD vacuum energy $V(\theta)$ at $\theta = 0$ \cite{Kawaguchi:2020kdl},
 
\begin{equation}
\chi_{\text {top }}=-\left.\int_T d^4 x \frac{\delta^2 V(\theta)}{\delta \theta(x) \delta \theta(0)}\right|_{\theta=0}
\end{equation}

$V(\theta)$ represents the effective potential of QCD, which includes the QCD $\theta$-term represented by the flavor-singlet gluonic operator. Since adding the $V(\theta)$ in holographic QCD is not straightforward, we will use the final form of the topological susceptibility  expressed by fermionic operators as \cite{Kawaguchi:2023olk}

\begin{equation}
\chi_{\text {top }}=-\frac{1}{4}\left[m_l\langle\bar{\psi} \psi\rangle+i m_l^2 \chi_{\eta_l} \right].
\end{equation}

By using Eq. \eqref{WI}, the topological susceptibility is written in terms of the pion susceptibility and light quark components of the $\eta$ susceptibility, and it is proportional to the square of light quark mass,

\begin{equation}
\chi_{\text {top }}=\frac{i m_l^{2}}{4}\left[\chi_{\pi}- \chi_{\eta_l} \right].
\label{chitop}
\end{equation}

To facilitate an understanding of the role of topological susceptibility, we
insert the scalar meson susceptibilities $\chi_{\sigma}$ and $\chi_{a_0}$ in Eq. \eqref{chitop},

\begin{equation}
\begin{aligned}
& \chi_{\text {top }}=\frac{i m_l^2}{4}\left[\left(\chi_\pi-\chi_\sigma\right)-\left(\chi_\eta-\chi_\sigma\right)\right] \\
& \chi_{\text {top }}=\frac{i m_l^2}{4}\left[\left(\chi_\pi-\chi_{a_0}\right)-\left(\chi_\eta-\chi_{a_0}\right)\right]
\end{aligned}
\end{equation}

Writing the $\chi_{\text {top }}$ in this form , one can realize that the
topological susceptibility is described by the combinations of the chiral partner $\chi_{\pi} \leftrightarrow \chi_{\sigma}$ ($\chi_{a_0} \leftrightarrow \chi_{\eta_{l}}$) and the $U(1)_A$ partner $\chi_{\pi} \leftrightarrow \chi_{a_{0}}$ ($\chi_{\eta_{l}} \leftrightarrow \chi_{\sigma}$). After the chiral restoration, the topological susceptibility is dominated by the $U(1)_A$ 
partner: $\chi_{\text {top }} \sim \chi_{\eta_{l}} - \chi_{\sigma}$ ($\chi_{\text {top }} \sim \chi_{\pi} - \chi_{a_{0}}$), so that $\chi_{\text {top }}$ acts as the indicator for the breaking strength of $U(1)_A$ symmetry. It should be noted that $\chi_{\text {top }}$ trivially vanishes in the
chiral limit ($m_l = 0$). Similarly, the Dilute Instanton Gas Approximation (DIGA) predicted the vanishing of the indicator of the breaking of $U(1)_A$ symmetry, $\chi_{\pi} - \chi_{a_{0}}$, for $N_f \ge 3$, such that the $\chi_{\pi} - \chi_{a_{0}} \propto m_l^{N_f -2 }$ \cite{Kovacs:2023vzi}.


\section{Results}
\label{sectionV}
The objective of this work is to examine the restoration of chiral symmetry and $U(1)_A$ symmetry at finite temperature by analyzing the screening mass and meson susceptibility using the soft-wall holographic QCD model. As mentioned earlier, we consider two cases with different forms for the 5D mass and dilaton field. The parameters for cases I and II can be found in Table \ref{tab:freeparcaseI} and Table \ref{tab:freeparcaseII}, respectively. These parameters are chosen to achieve a physical pion mass and a pseudocritical temperature of approximately $155$ MeV for both cases.

\begin{figure} 
  \centering
  \includegraphics[width=0.49\linewidth]{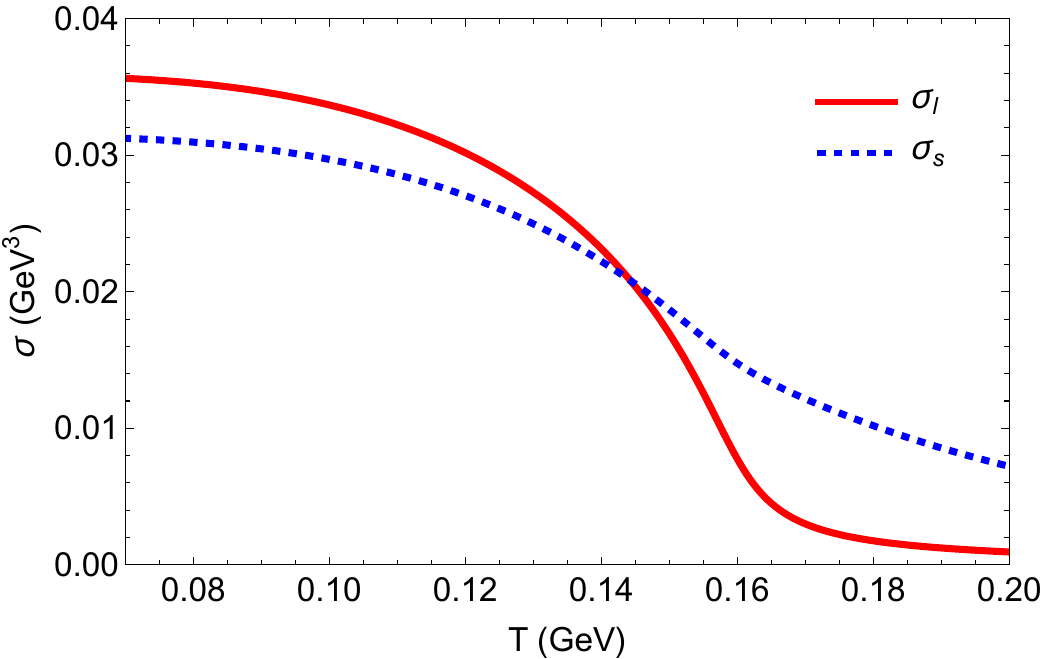}   
  \includegraphics[width=0.49\linewidth]{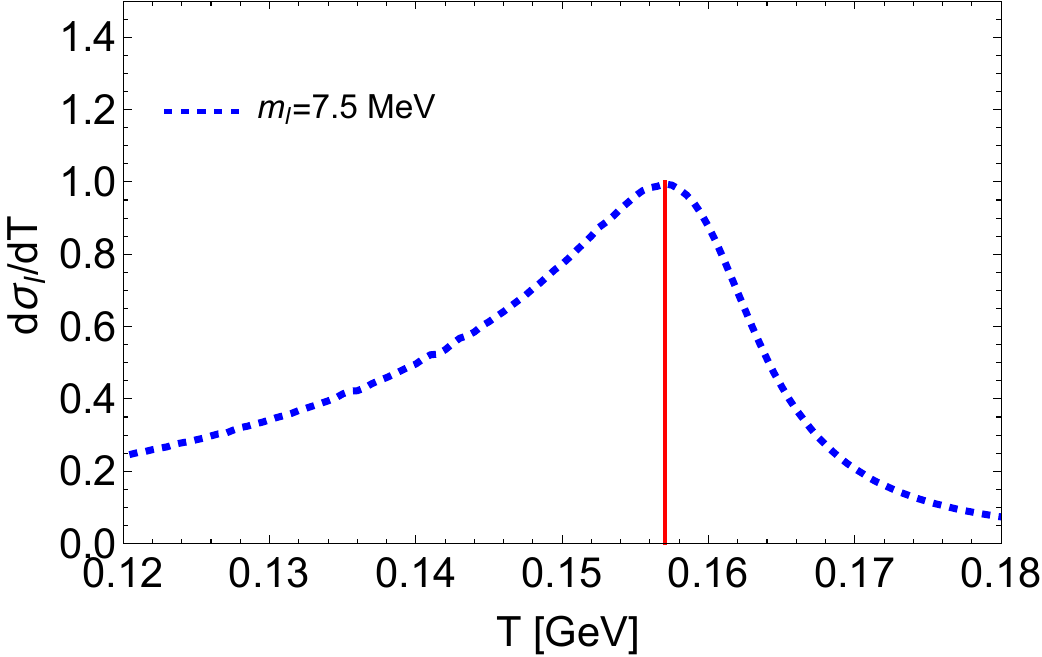} 

\caption{Left: The light quark condensate $\sigma_l$ and the strange quark condensate $\sigma_s$ as functions of temperature $T$.
Right: Evaluating the pseudocritical temperature at $T_{pc}=0.157$ GeV from the peak of $-d \sigma_l/dT$. 
These results are for case I with $m_l=7.5\,{ \rm MeV}$.
}
\label{condensateI}
\end{figure}

\begin{figure} 
  \centering
  \includegraphics[width=0.49\linewidth]{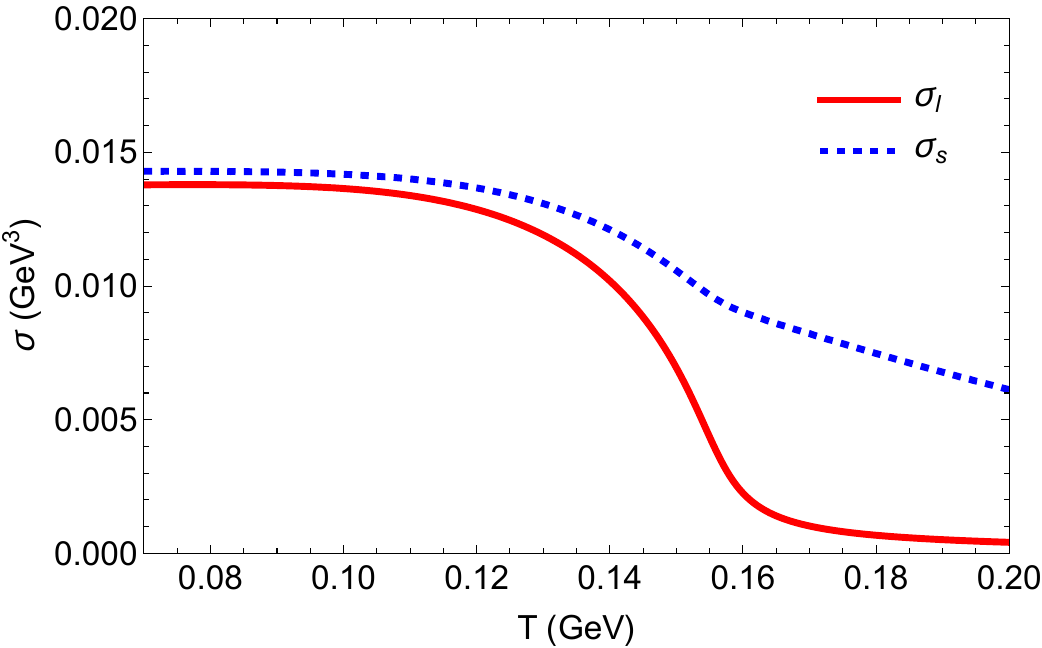}   
  \includegraphics[width=0.49\linewidth]{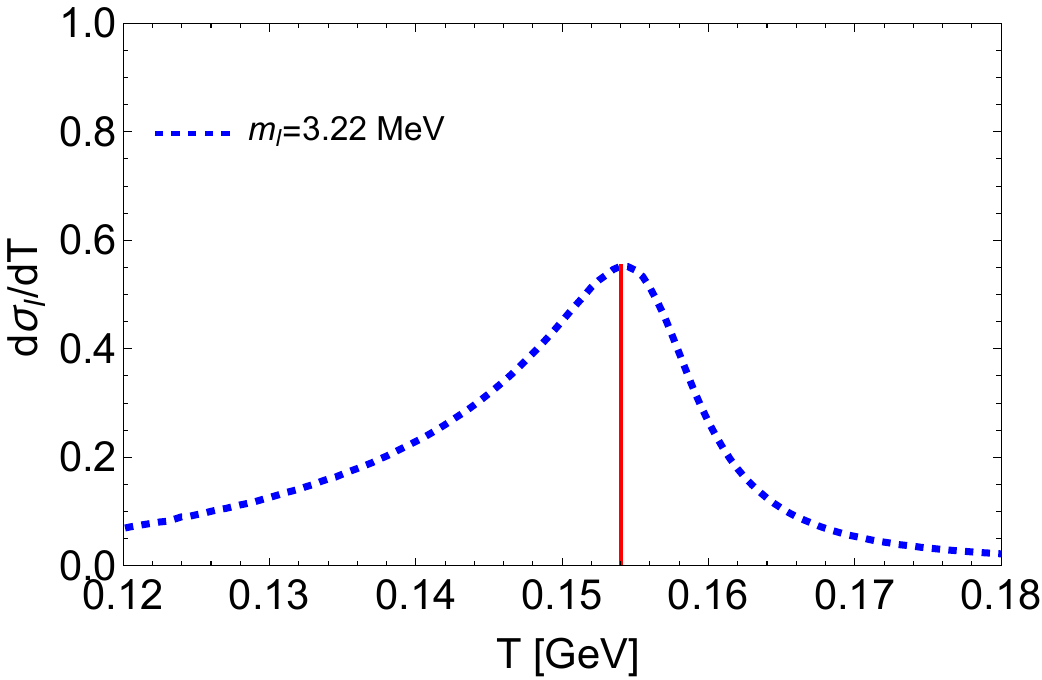} 

\caption{The color online is similar to Fig. \ref{condensateI}. The evaluated pseudocritical temperature is $T_{pc}=0.154$ GeV with the case II parameter setup.}
\label{condensateII}
\end{figure}

First, we discuss the chiral phase transition at physical quark masses with three-quark flavors. In Fig.~\ref{condensateI}, we present the light quark condensate $\sigma_{l}$ and $\sigma_{s}$ as functions of temperature for the three-quark flavor system with physical quark masses in case I. The left panel demonstrates that the light chiral condensate $\sigma_{l}$ decreases smoothly with increasing temperature. Because the finite quark mass ($m_l \neq 0$) breaks chiral symmetry explicitly, the transition is a smooth crossover rather than a sharp drop. Above the pseudo-critical temperature, $\sigma_{l}$ retains a small, finite value, but it continues to melt away steadily toward zero at higher temperatures. This confirms that chiral symmetry is effectively restored in the high-temperature phase, consistent with standard lattice QCD and effective model results.
. The chiral crossover is characterized by the pseudocritical temperature, which is determined by the inflection point of the quark condensate, $d^2 \sigma_l/d^2 T \bigl|_{T=T_{\rm pc}}=0$, and is evident as a peak in $- d \sigma_l/dT$ as shown in the right panel of Fig.~\ref{condensateI}. The strange quark condensate exhibits similar behavior, with a higher value at high temperatures compared to the light quark condensate. For the three-quark flavor system with physical masses of light and strange quarks, the pseudocritical temperature is determined to be $T_{\rm pc}\bigl|_{\rm hQCD} = 0.157~{\rm GeV}$. A similar result for case II is shown in Fig.~\ref{condensateII}, where the pseudocritical temperature is determined to be $T_{\rm pc}\bigl|_{\rm hQCD} = 0.154~{\rm GeV}$.

\begin{table} 
\center
\begin{tabular}{c c  c cc c c}
\hline
\hline

  $m_l=7.5$ MeV ($m_{\pi}=132.5$ MeV) &    &  $m_l=13$ MeV ($m_{\pi}=174.4$ MeV)  &  &$m_s=95$ MeV &    &  $\mu_{g} = 0.44$ (GeV)       \\ 
  $\mu_{1} =0.81$ (GeV)      &      & $\mu_{2} =0.176$ (GeV)      &      &  $\lambda = 80$  &      &   $\gamma = -10$\\
  
     \hline
     \hline
\end{tabular}
\caption{The values of the free parameters for case I.}
\label{tab:freeparcaseI}
\end{table}

\begin{table} 
\center
\begin{tabular}{c c  c cc c c}
\hline
\hline

  $m_l=3.22$ MeV ($m_{\pi}=141.6$ MeV) &    &  $m_l=6$ MeV ($m_{\pi}=193$ MeV)  &  &$m_s=95$ MeV &    &  $\mu_{g} = 0.44$ (GeV)       \\ 
  $\mu_{c} =1.27$ (GeV)      &      &   $\lambda = 80$  &      &   $\gamma = -23$\\
  
     \hline
     \hline
\end{tabular}
\caption{The values of the free parameters for case II.}
\label{tab:freeparcaseII}
\end{table}

\subsection{Screening mass}

In a vacuum, mesons are characterized by their masses and decay widths in different channels. At finite temperature, the concept of mass is replaced by screening mass and pole mass due to the breaking of Lorentz symmetry. Both of these masses carry information about the correlation function of the meson field in a hot medium. The screening mass, which can be evaluated as the pole of the spatial component of the two-point correlation function, serves as an indication of chiral symmetry restoration. This is because the mass difference of chiral partner mesons ($\pi$ and $\sigma$, and $\eta$ and $a_0$) originates from the spontaneous chiral symmetry breaking of QCD. After the chiral symmetry restoration, the masses of chiral partners should be degenerate.

In our work, we calculated the two-point Green's function for $a_0$, $\sigma$, $\pi$, and $\eta$ fields. The screening mass of the $a_0$ meson is obtained as the pole of $G_{a_0}$ in Eq. \eqref{Ga0}, similar to the condition of $s_1 (p^2)=0$, which corresponds to the pole of $G_{a_0}$. The smallest spatial value of $p^2$ satisfying the condition $s_1 (p^2)=0$ is equal to the screening mass of $a_0$, $M_{a_0,scr}^{2}=-p^2$. Similarly, we can find the screening mass of the $\sigma$ meson.

\begin{figure} 
  \centering
  \includegraphics[width=0.49\linewidth]{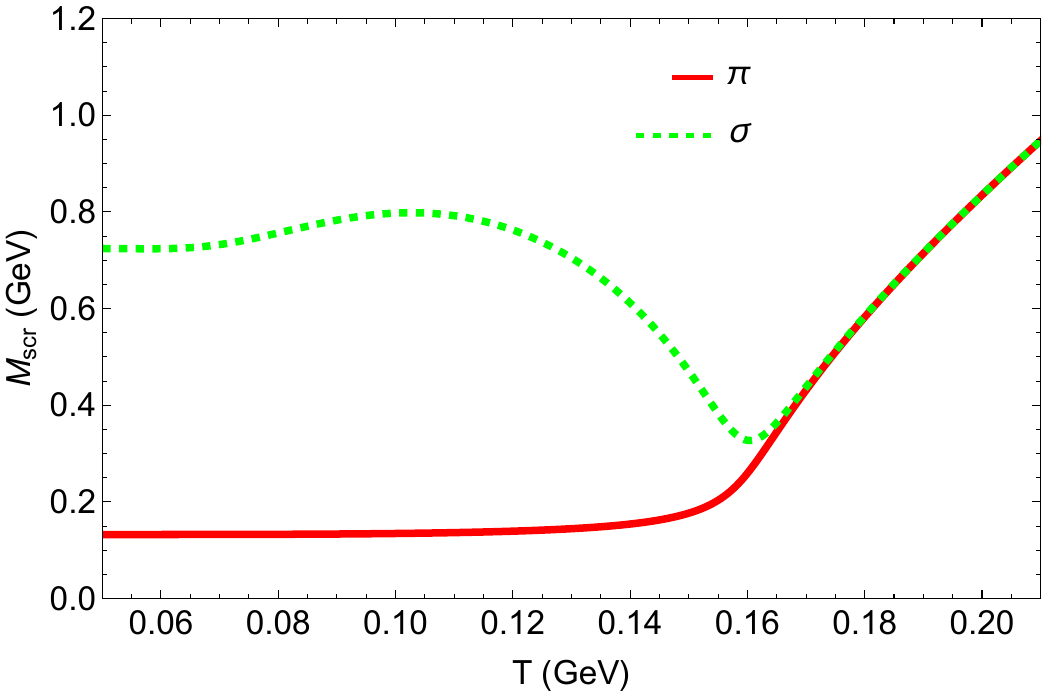}   
  \includegraphics[width=0.49\linewidth]{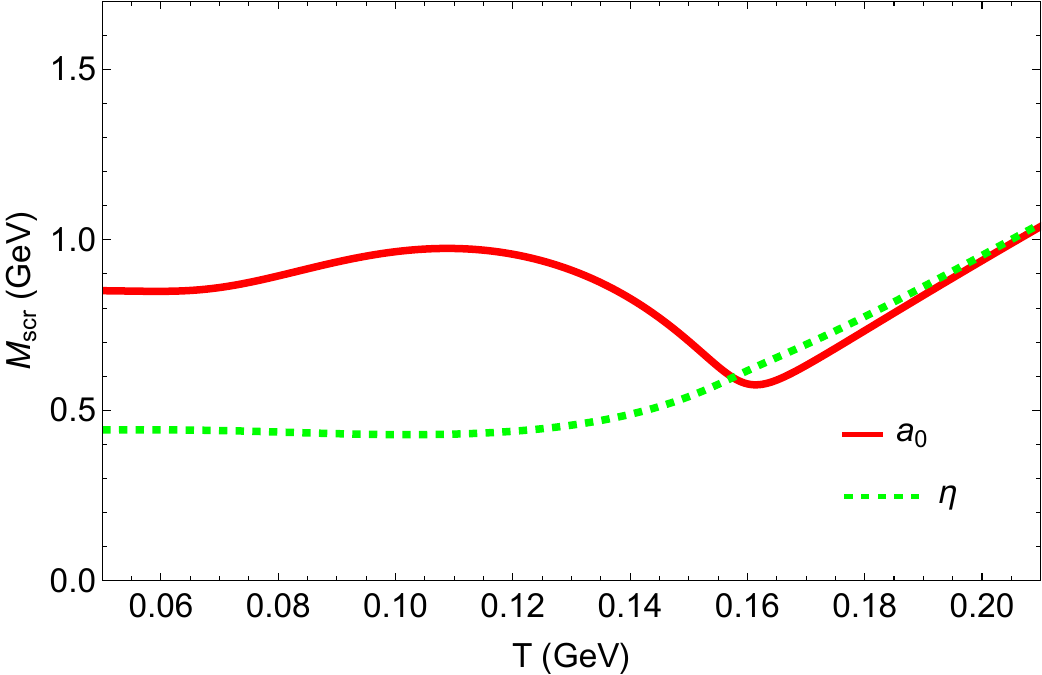} 

\caption{Screening mass of the chiral symmetry partners ($\pi$ and $\sigma$, and $\eta$ and $a_0$) as a function of temperature for case I.}
\label{screen}
\end{figure}

\begin{figure} 
  \centering
  \includegraphics[width=0.49\linewidth]{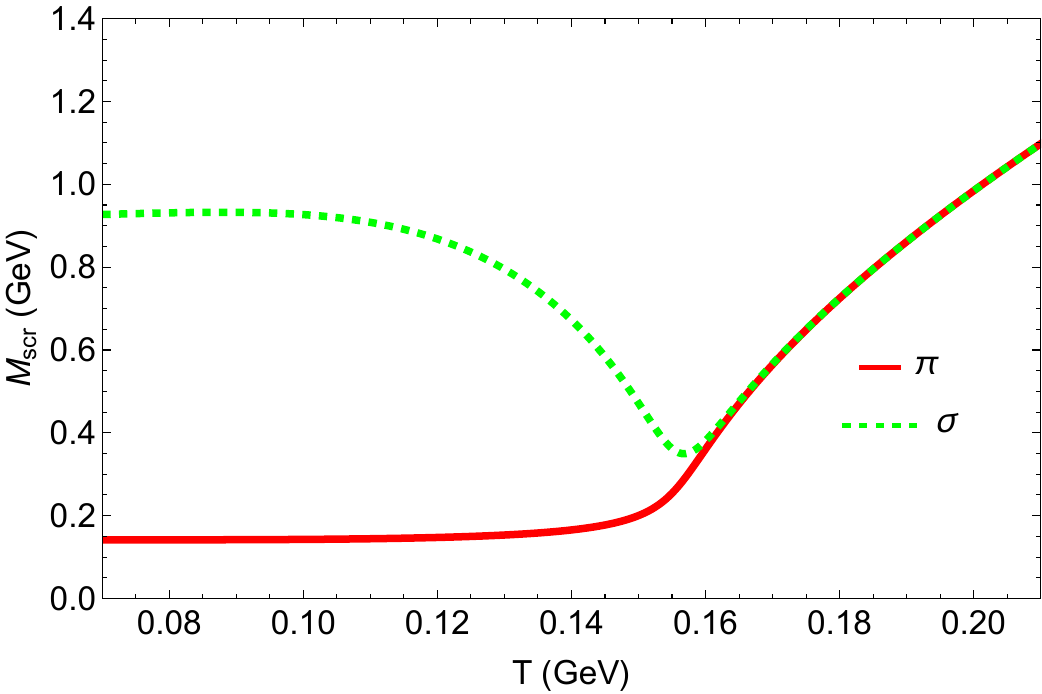}   
  \includegraphics[width=0.49\linewidth]{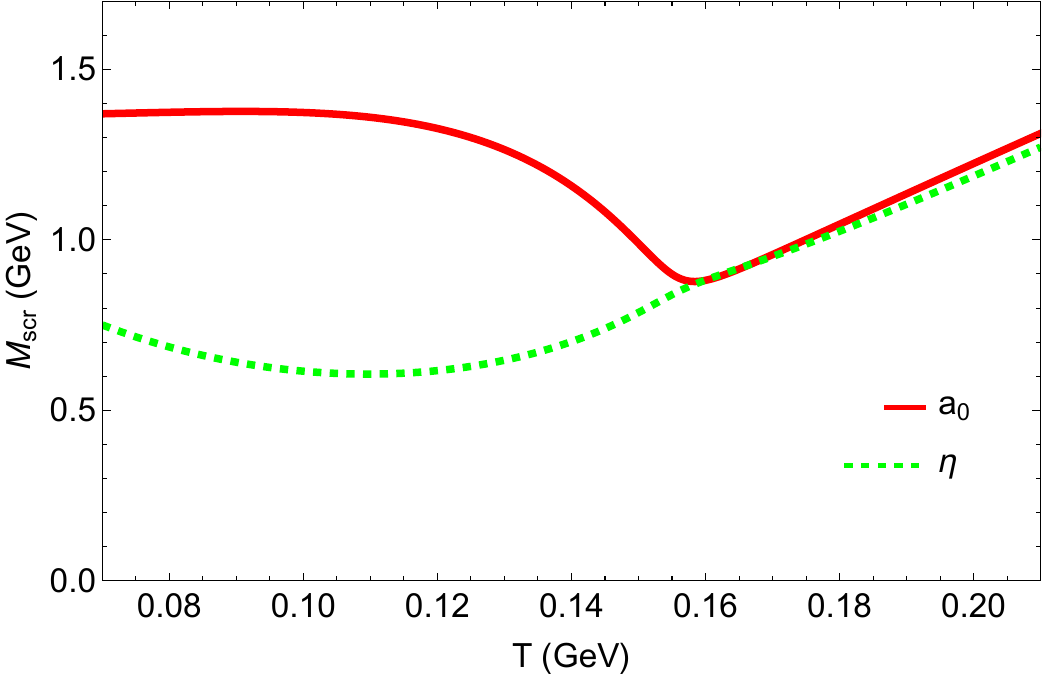} 

\caption{The color online is similar to Fig. \ref{screen} for case II.}
\label{screenII}
\end{figure}

For the pseudoscalar sector, we need to follow the same method and find the pole of the two-point Green's functions, which correspond to the conditions $\pi_0 (p^2)=0$ and $\pi_{\eta,0} (p^2)=0$ for the screening mass of the pion and $\eta$, respectively. The results of the screening masses of the chiral partners are shown in Fig. \ref{screen} and Fig. \ref{screenII} for case I and case II, respectively. From these figures, we can see that near the pseudocritical temperature, the masses of chiral partners become degenerate, providing a signal of the chiral symmetry restoration. Quantitatively, these results are consistent with the ones present from LQCD \cite{Cheng:2010fe} and Nambu-Jona-Lasinio (NJL) \cite{Jiang:2011aw}, and confirm that the temperature-dependent behavior of screening masses of the chiral partner mesons is strongly coupled with the chiral phase transition.

\subsection{Meson susceptibility}

In the study of symmetry restoration (chiral symmetry and $U(1)_A$ symmetry), an important indicator is the meson susceptibility, which measures the response of the system to the addition of a field. As demonstrated in the previous section, the pion susceptibility requires a normalization factor to align with the one obtained from the WTI. Similarly, for the $\sigma$ susceptibility, a direct method can be used to determine $\chi_{\sigma}$ by taking the derivative of the chiral condensate with respect to the light quark mass:

\begin{equation}
    \chi_{\sigma}= \frac{\partial \sigma_l}{\partial m_l}.
\label{chisig2}
\end{equation}

To relate the present definition of the susceptibility to the one given in Eq.~\eqref{chisig}, note that the two expressions differ only by an overall normalisation factor $\zeta^2/4$. A small deviation to the quark mass $m_l+\delta m_l$ will cause a small deviation $\delta\chi_l$ to $\chi_l$. Obviously, the linearised fluctuation $\delta \chi$ obeys the same equation of motion as Eq.~\eqref{pscalareq} evaluated at zero four-momentum, $p^2=0$. Therefore, the UV expansion of $\delta\chi_l$ should be $\delta\chi_l=s_1 z+...+s_3 z^3+...$. On the other hand, the dual field-theory interpretation identifies the leading and sub-leading coefficients with the quark-mass and condensate variations: $s_1= \delta m_l \zeta$ and $s_3=\delta \sigma/\zeta$, and the susceptibility defined in Eq.~\eqref{chisig} is therefore $\chi_\sigma\sim 4 s_3/s_1=4(\delta \sigma/\zeta)/(\delta m_l \zeta)=\frac{4}{\zeta^2} \frac{\partial \sigma}{\partial m_l}$.

\begin{figure} 
  \centering
  \includegraphics[width=0.49\linewidth]{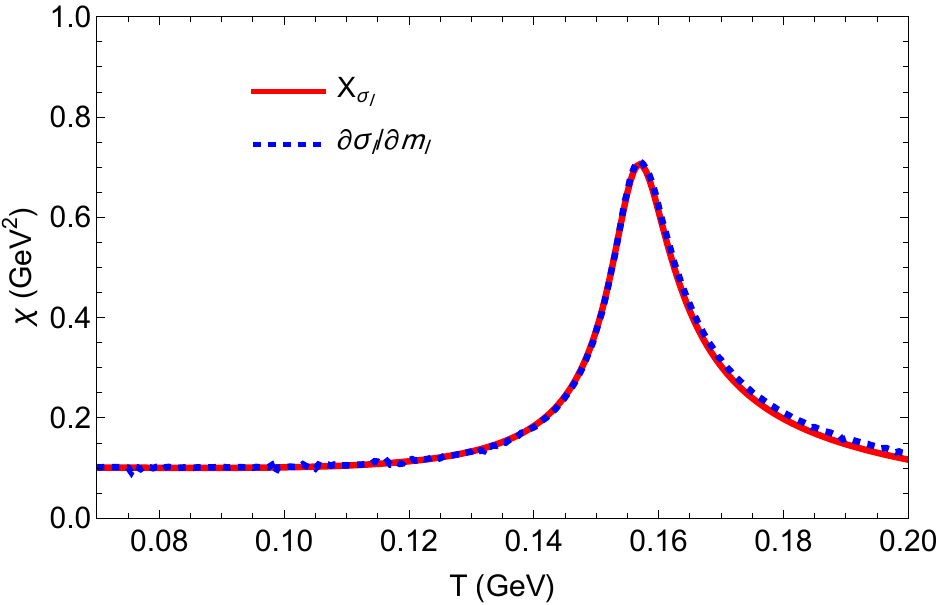}   

\caption{Comparison of $\chi_{\sigma}$ in Eq. \eqref{chisig2} with the normalized $\chi_{\sigma}$ in Eq. \eqref{chisig}, with the normalization factor $N_{\sigma}=\zeta^2/4$.}
\label{comp}
\end{figure}

In Fig. \ref{comp}, we compare the numerical results of $\chi_{\sigma}$ in Eq. \eqref{chisig2} with the normalized $\chi_{\sigma}$ in Eq. \eqref{chisig}. The normalization factor, found by equating Eq. \eqref{chisig2} and Eq. \eqref{chisig}, is $N_{\sigma}=\zeta^2/4$, which aligned with the normalization factor we obtained analytically. 

Similar to the concept of screening mass, when two mesons are symmetry partners, their susceptibilities should be equal after symmetry restoration. The sign of chiral symmetry restoration can be observed in the following conditions.

\begin{equation}
    \begin{gathered}
     \chi_{\pi} - \chi_{\sigma} \to 0 ~,\\
      \chi_{\eta_l} - \chi_{a_0}  \to 0 ~.
    \end{gathered}
\end{equation}

Similarly, the indication of $U(1)_A$ symmetry restoration can be read from 

\begin{equation}
    \begin{gathered}
     \chi_{\pi} - \chi_{a_0} \to 0 ~,\\
      \chi_{\eta_l} - \chi_{\sigma} \to 0 ~.
    \end{gathered}
\end{equation}

\begin{figure} 
  \centering
  \includegraphics[width=0.49\linewidth]{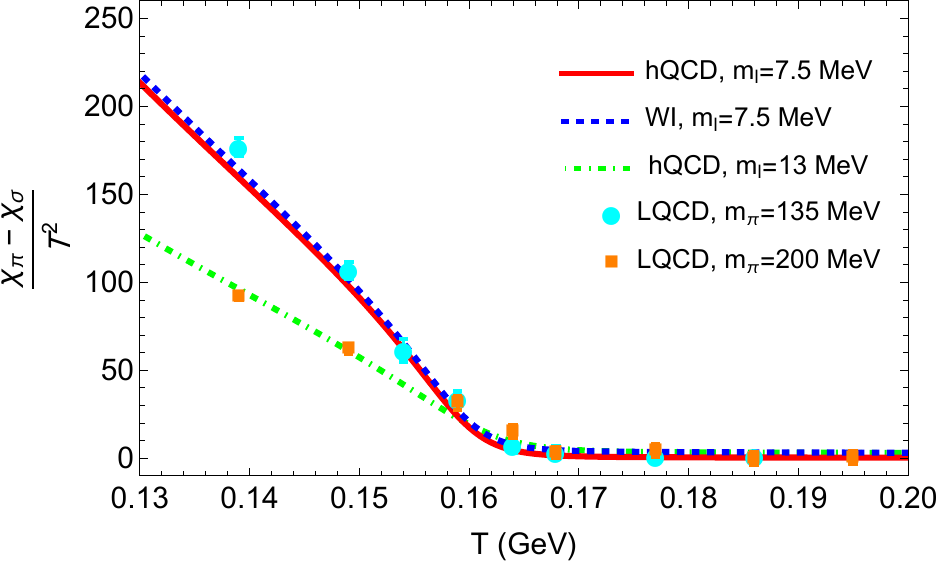}   
  \includegraphics[width=0.47\linewidth]{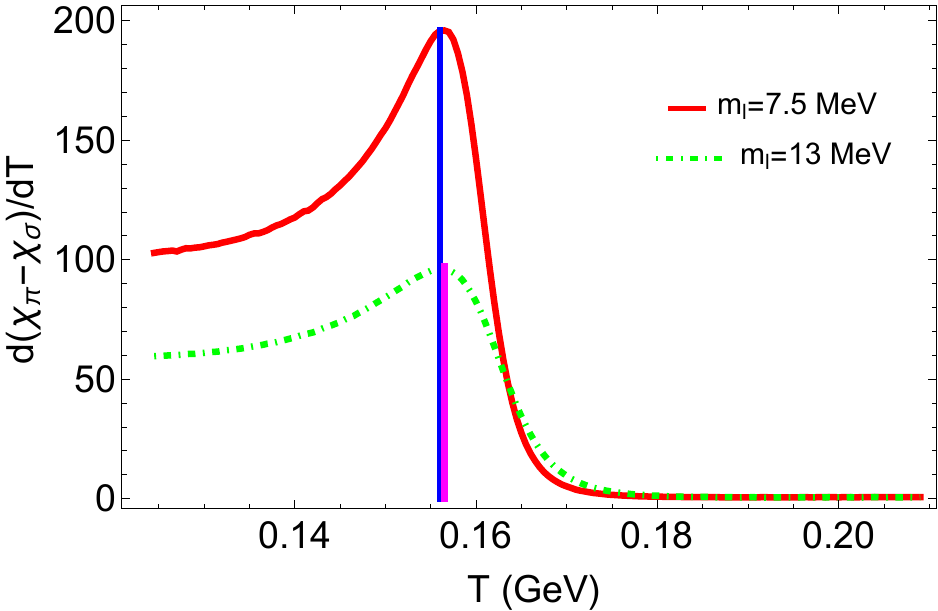}
  
  \includegraphics[width=0.49\linewidth]{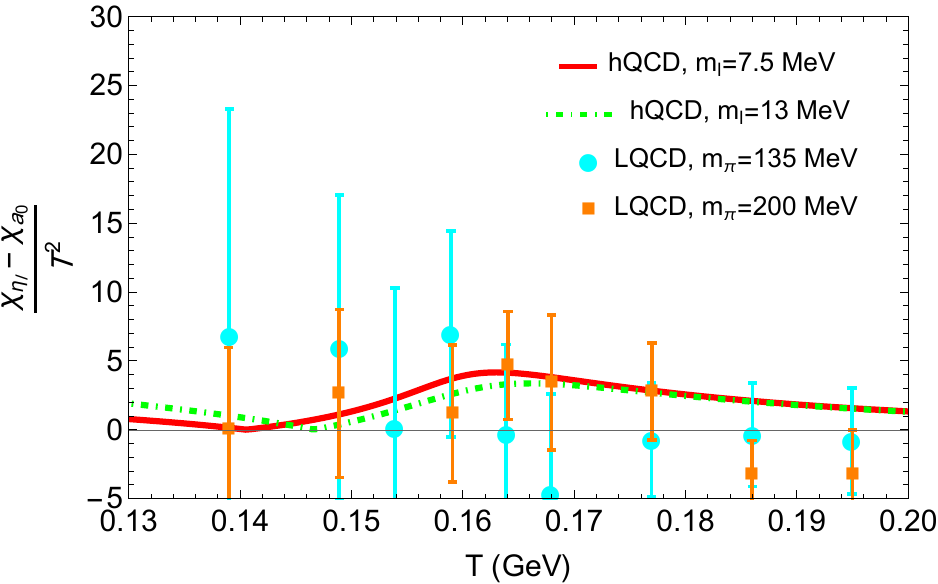} 

\caption{ The indicator of the chiral symmetry restoration $( \chi_{\pi} - \chi_{\sigma})/T^2$ (top-left) and $(\chi_{\eta_l} - \chi_{a_0})/T^2$ (bottom) as a function of temperature at $m_l=7.5$ MeV and $m_l=13$ MeV for case I. The LQCD data is taken from Ref. \cite{Bhattacharya:2014ara}.}
\label{chisu2I}
\end{figure}

\begin{figure} 
  \centering
  \includegraphics[width=0.49\linewidth]{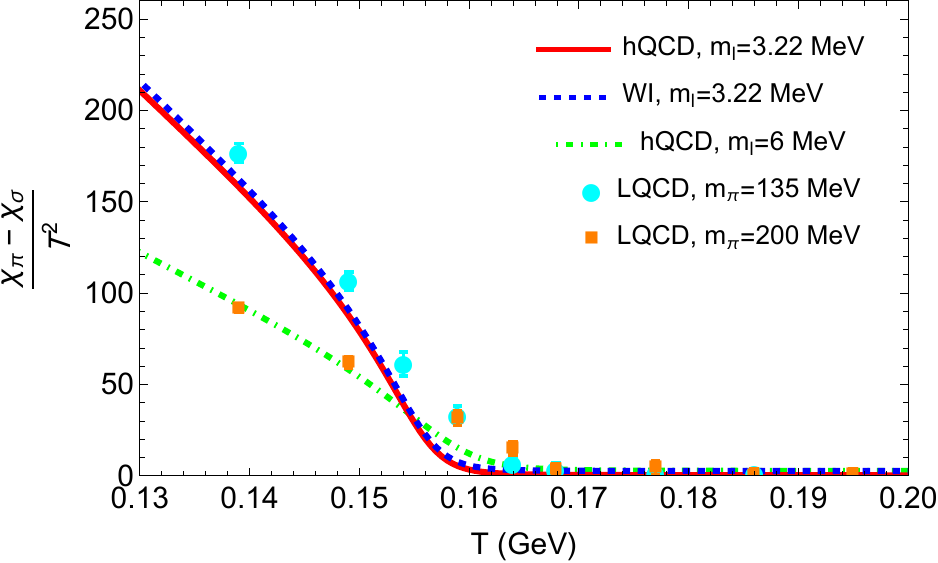}  
  \includegraphics[width=0.47\linewidth]{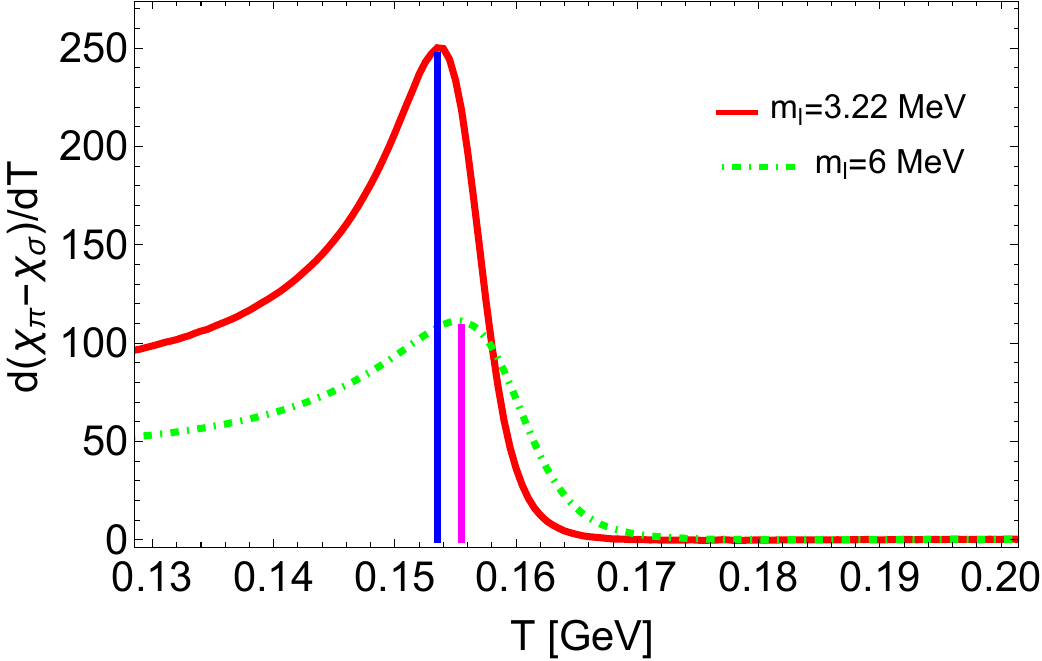}
  
  \includegraphics[width=0.49\linewidth]{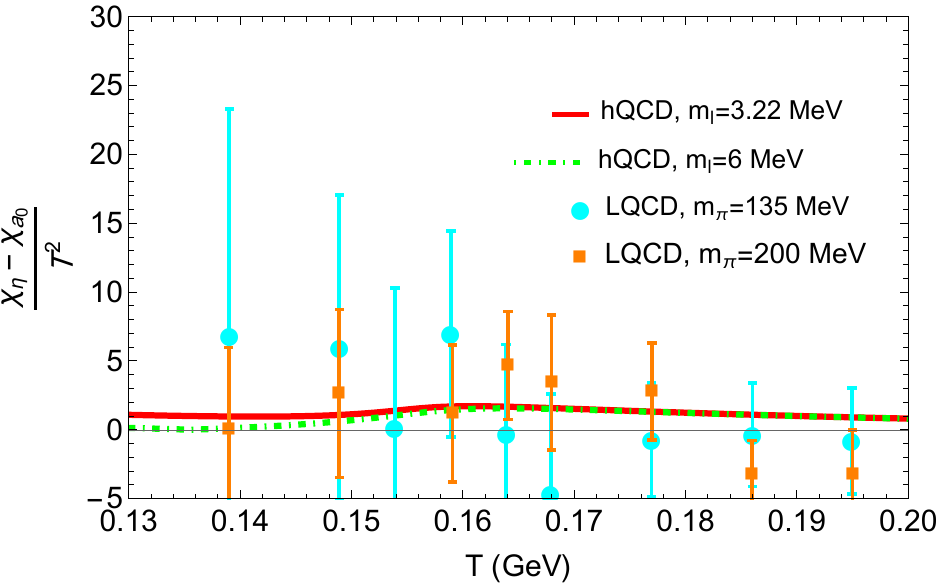} 

\caption{ The color online is similar to Fig. \ref{chisu2I} with the parameterization of case II}
\label{chisu2II}
\end{figure}

The results of the $\chi_{\pi} - \chi_{\sigma}$ and $\chi_{\eta_l} - \chi_{a_0}$ as a function of temperature are shown in Fig. \ref{chisu2I} and Fig. \ref{chisu2II} for case I and case II, respectively.
As shown in the top-left panel of Fig. \ref{chisu2I}, the indicator ($\chi_{\pi} - \chi_{\sigma}$) for the isoscalar partners ($\sigma$, $\pi$)  drops sharply near the crossover temperature, signaling the rapid restoration of chiral symmetry. This behavior is observed for both light quark masses $m_l=7.5$ MeV and $m_l=13$ MeV in our model and is consistent with the observation in lattice LQCD data at physical quark masses \cite{Bhattacharya:2014ara}. The pseudocritical temperature estimated, from the top-right panel of Fig. \ref{chisu2I}, is similar to the one found from Fig. \ref{condensateI}, confirming the restoration of the chiral symmetry from the meson susceptibility. 

The bottom panel of Fig. \ref{chisu2I} displays the susceptibility difference of isovector partners $a_0$ and $\eta_l$ at light quark masses $m_l=7.5$ MeV and $m_l=13$ MeV. Despite the higher restoration point compared to the isoscalar partners, the results are compatible with the LQCD data within the uncertainty. The discrepancy in the restoration temperatures between the two chiral partner pairs is a direct signature of the influential role played by the $U(1)_A$ anomaly. The $\eta$ meson receives a significant mass contribution from the anomaly, which is reflected in its susceptibility. Therefore, for the susceptibilities of the isovector pair to become degenerate, it is not sufficient for the chiral condensate to melt; the effects of the $U(1)_A$ anomaly must also be sufficiently suppressed.

For case II (Fig. \ref{chisu2II}), the behavior of $\chi_{\pi} - \chi_{\sigma}$ qualitatively agrees with LQCD, and $\chi_{\eta_l} - \chi_{a_0}$ is compatible with LQCD for both $m_l=3.22$ MeV and $m_l=6$ MeV. The pseudocritical temperature evaluated for case II is also consistent with Fig. \ref{condensateII}.

\begin{figure} 
  \centering
  \includegraphics[width=0.55\linewidth]{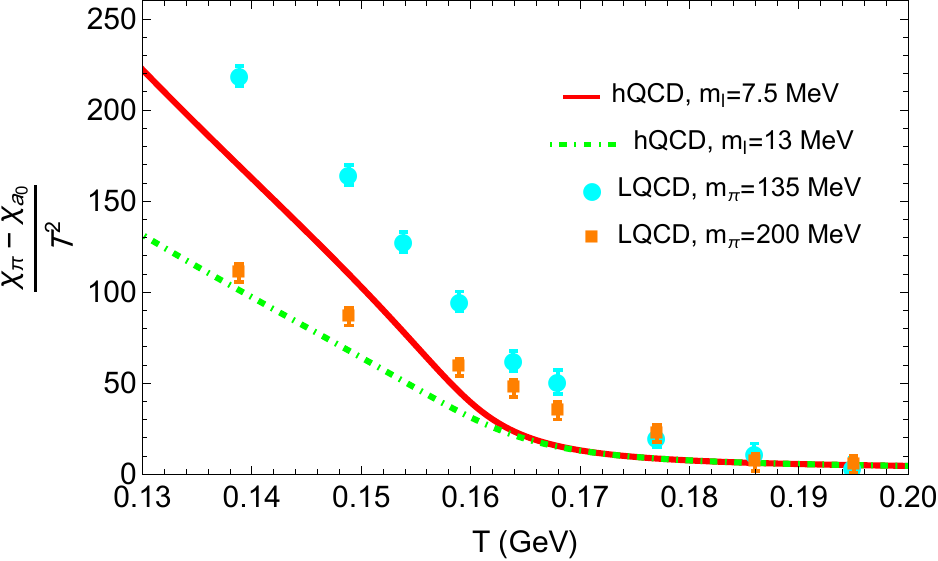}   

\caption{ The indicator of $U(1)_A$ symmetry breaking as a function of temperature at $m_l=7.5$ MeV and $m_l=13$ MeV. The susceptibilities difference, $\chi_{\pi}-\chi_{a_0}$, becomes quark mass independent for $T > 0.165$ GeV. These results are for case I.  The LQCD data is taken from Ref. \cite{Bhattacharya:2014ara}. }
\label{chiU2I}
\end{figure}

\begin{figure} 
  \centering
  \includegraphics[width=0.55\linewidth]{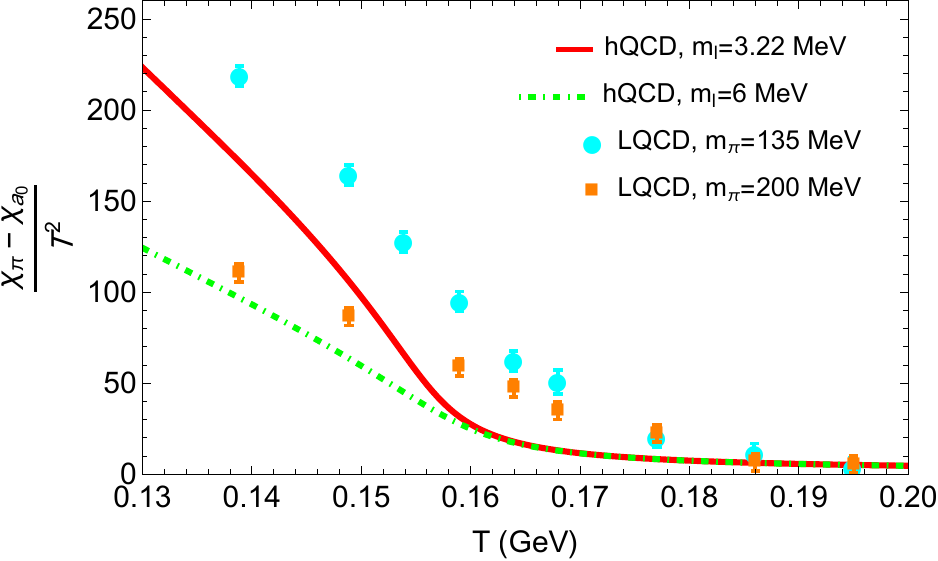}   

\caption{ The color online is similar to Fig. \ref{chiU2I} with case II parameterization.}
\label{chiU2II}
\end{figure}

In order to identify the breaking of the $U(1)_A$ symmetry, we analyze the difference between the susceptibilities of the $U(1)_A$ symmetry partners $\pi$ and $a_0$ ($\chi_{\pi}-\chi_{a_0}$).
Although there is no strict definition for the exact $U(1)_A$ restoration, particularly since the estimators do not strictly vanish for non-zero quark mass \cite{Giordano:2025shr}. In our analysis, we evaluate the magnitude of $U(1)_A$ restoration via the significant suppression of the susceptibility difference between the $U(1)_A$ partners, as displayed in Fig. \ref{chiU2I} and Fig. \ref{chiU2II} for cases I and II, respectively. Unlike the results for chiral symmetry restoration, the outcomes in  Fig. \ref{chiU2I} do not align with the LQCD data at the low and mid temperatures. However, the restoration temperature of the $U(1)_A$ symmetry is aligned with LQCD, and is slower than the restoration of the chiral symmetry. The degeneracy of the results for $m_l=7.5$ MeV and $m_l=13$ MeV for $T > 0.165$ GeV proves that the non-zero value of the $\chi_{\pi}-\chi_{a_0}$ comes from  the axial anomaly, not the small quark mass.  This indicates that, within our current holographic construction, the dynamics driving the chiral transition are different from those controlling the axial anomaly strength, which signals the emergence of two distinct energy scales.

The situation described for Case I is found to be robust and is replicated in the alternative parameter set, Case II. As shown in Fig. \ref{chiU2II}, the $U(1)_A$ breaking indicator for Case II also undergoes its most rapid change at a temperature that coincides with the chiral pseudo-critical temperature for this situation. The restoration temperature, similar to case I, is slower than chiral symmetry. Then, within the holographic QCD, the restoration of the $U(1)_A$ axial symmetry is independent of the specific parameter tuning within the studied range.

Historically, the exact relationship between the restoration of the chiral symmetry and the anomalous $U(1)_A$ axial symmetry at finite temperature has been a subject of intense debate \cite{Gross:1980br,Pisarski:1980md,Schafer:1996wv,Shuryak:1993ee}. In this work, our soft-wall holographic model yields a clear separation of scales, where the effective restoration of $U(1)_A$ occurs at a temperature significantly higher than the pseudo-critical chiral crossover temperature.
While earlier conventions frequently assumed a simultaneous restoration of both symmetries at $T_c$, our qualitative feature of scale separation aligns remarkably well with a vast body of recent literature. On the lattice side, independent formulations using highly anisotropic lattices by the Fastsum collaboration \cite{Bignell:2026ybw,Aarts:2026kpq} report $T_{U(1)_A} \sim 320\text{ MeV}$, well above their computed chiral crossover scale ($T_{pc} \sim 154 - 180\text{ MeV}$). Similarly, early and recent Highly Improved Staggered Quark (HISQ) and domain-wall computations by the HotQCD collaboration \cite{Bhattacharya:2014ara,HotQCD:2012vvd} confirm that while $SU(2)_L \times SU(2)_R$ is restored swiftly above $T_c \sim 155\text{ MeV}$, anomalous $U(1)_A$ breaking observables remain distinctly non-zero and persist deep into the chirally symmetric phase \cite{Kaczmarek:2021ser,Ding:2021jtn}. This behavior is further supported by functional renormalization group (FRG) \cite{Li:2019chs} and Polyakov-loop quark-meson (PQM) \cite{Rai:2018ufz} frameworks, where a temperature-dependent 't Hooft determinant coupling drives a distinct lag between the melting of the light quark condensates and the full restoration of the axial anomaly. Furthermore, analyses of symmetry breaking strengths in different channels suggest a two-stage restoration scenario \cite{Chiu:2026sxy}. In this picture, chiral symmetry and $U(1)_A$ symmetry are first restored in the nonsinglet sector near $T_c$, but full restoration, including the singlet channels, occurs at a much higher temperature when topological fluctuations are largely suppressed.

Regarding the universality class of the chiral transition: lattice QCD simulations with physical quark masses indicate a smooth crossover rather than a genuine phase transition \cite{Buchoff:2013nra}. In the chiral limit, the transition is expected to belong to the 3D O(4) universality class for two light flavors, with the strange quark acting as a heavy spectator that does not alter the critical exponents \cite{Kaczmarek:2021ser}. In our holographic framework, as we have shown in detail in our previous work \cite{Ahmed:2024rbj}, for the $2+1$ flavor case, the order of the chiral phase transition is a smooth crossover at the physical quark masses. However, moving to the chiral limit, the order of the phase transition changes from crossover to first order due to the contribution of $U(1)_A$ Anomaly. Moreover, in a case similar to the LQCD, with massless light quarks and the physical strange quark mass, the order of the phase transition shows a second-order line similar to the 3D O(4) universality class. However, because classical gravity duals solve the field theory in the mean-field approximation, our model yields standard mean-field critical exponents rather than true 3D $O(4)$ scaling \cite{Ahmed:2024rbj}. Incorporating $1/N_c$ loop corrections to the bulk gravitational action would be required to capture the full non-perturbative thermodynamic fluctuations necessary to resolve these universality benchmarks cleanly.

\subsection{Topological susceptibility}

\begin{figure} 
  \centering
  \includegraphics[width=0.49\linewidth]{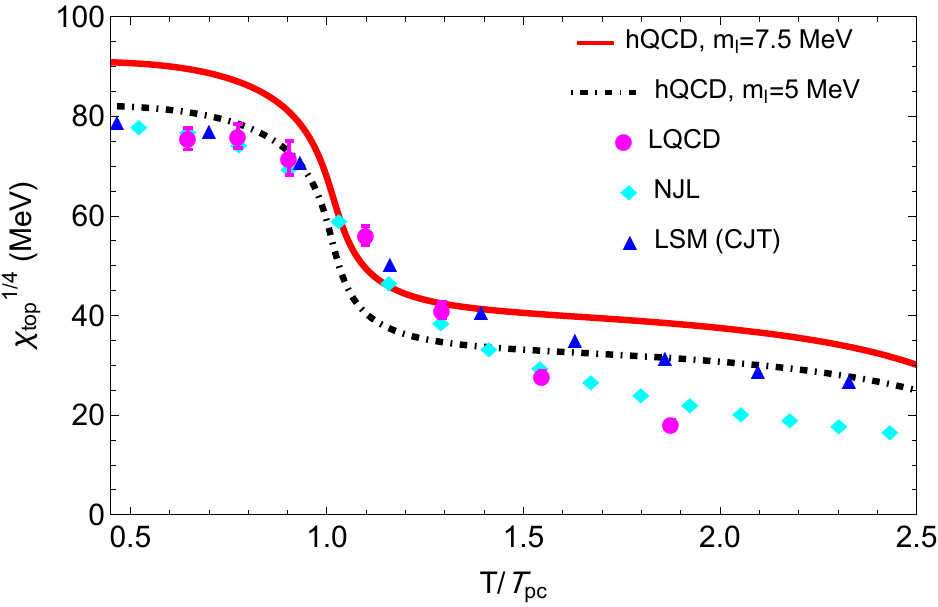}   
  \includegraphics[width=0.49\linewidth]{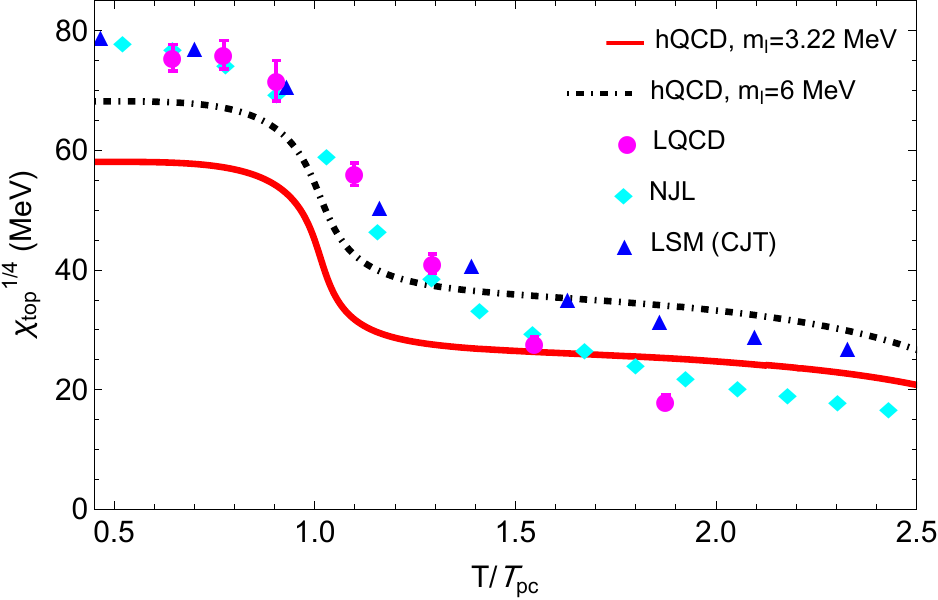}   
\caption{ The topological susceptibilities as a function of temperature at different light quark masses for case I (left) and II (right). The other approaches are LQCD \cite{Borsanyi:2016ksw}, NJL \cite{Cui:2021bqf}, and LSM (CJT) \cite{Kawaguchi:2020qvg}. }
\label{topsusI}
\end{figure}

Using the WTI associated with chiral symmetry, it can be demonstrated that the topological susceptibility is linked to the chiral- and $U(1)_A$ partner structures in the meson susceptibility functions. Therefore, the topological susceptibility can also be considered as an indicator of the strength of $U(1)_A$ symmetry breaking during the chiral phase transition. We numerically evaluate $\chi_{\text{top}}$ in Eq.\eqref{chitop}, with the holographic QCD estimates on the $\chi_{\pi}$ and $\chi_{\eta_l}$
susceptibilities, as a function of temperature. In Fig. \ref{topsusI}, we illustrate the temperature dependence of the topological susceptibility $\chi_{\text{top}}^{1/4}$, where we have taken the absolute value of $\chi_{\text{top}}$, and compare it with LQCD data, NJL model, and linear sigma model (LSM). For case I (left panel), the value of $\chi_{\text{top}}^{1/4}(T=0)$ at the light quark mass $m_l=7.5$ MeV, which corresponds to the physical pion mass, is larger than the values obtained using other methods, possibly due to the fact that the physical quark mass used in other models is $m_l \sim 5$ MeV. Therefore, we also consider the case $m_l=5$ MeV and the $\chi_{\text{top}}^{1/4}(T=0)$ is in agreement with other models. However, in both cases of two light quark masses, the temperature behaviour is the same, and the $\chi_{\text{top}}^{1/4}$ declines more rapidly than others near the pseudocritical temperature. As we demonstrated in the previous subsection, in our model, the quantitative behavior of the indicator of $U(1)_A$ axial anomaly as a function of temperature is not consistent with LQCD, and it decreases sharply near the pseudocritical temperature. This feature causes the $\chi_{\text{top}}^{1/4}$ to decrease more sharply than in other models around the pseudocritical temperature. However, at higher temperatures our results show more persistent compare to other approaches. This slower suppression in our model points to specific features of the holographic gravitational dual. The dynamics of the axial anomaly and topological sectors are encoded in the structure of the bulk fields and their interactions. This could stem from the large-$N_c$ nature underlying holographic constructions, where certain topological effects are enhanced, or from the specific choice of the gravitational background that does not fully capture the deconfining dynamics of the gauge theory. Overall, we can conclude that the qualitative behaviour of our model is similar to other approaches, as the value of $\chi_{\text{top}}^{1/4}$ is almost constant for $T < T_{pc}$. Then, around the $T \sim T_{pc}$, the curve decreases sharply, and finally, it slowly decreases for $T > T_{pc}$. The quantitative behavior of $\chi_{\text{top}}^{1/4}$ for case II (right panel of Fig. \ref{topsusI}) is not better than case I in comparison with other approaches.

\begin{figure} 
  \centering
  \includegraphics[width=0.49\linewidth]{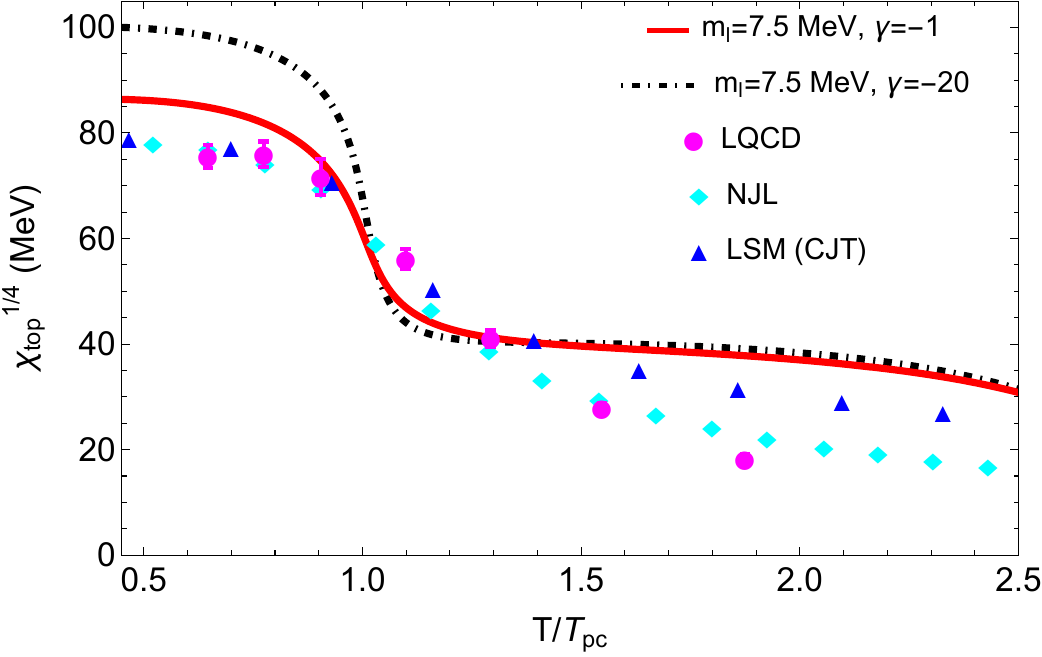}   
  \includegraphics[width=0.49\linewidth]{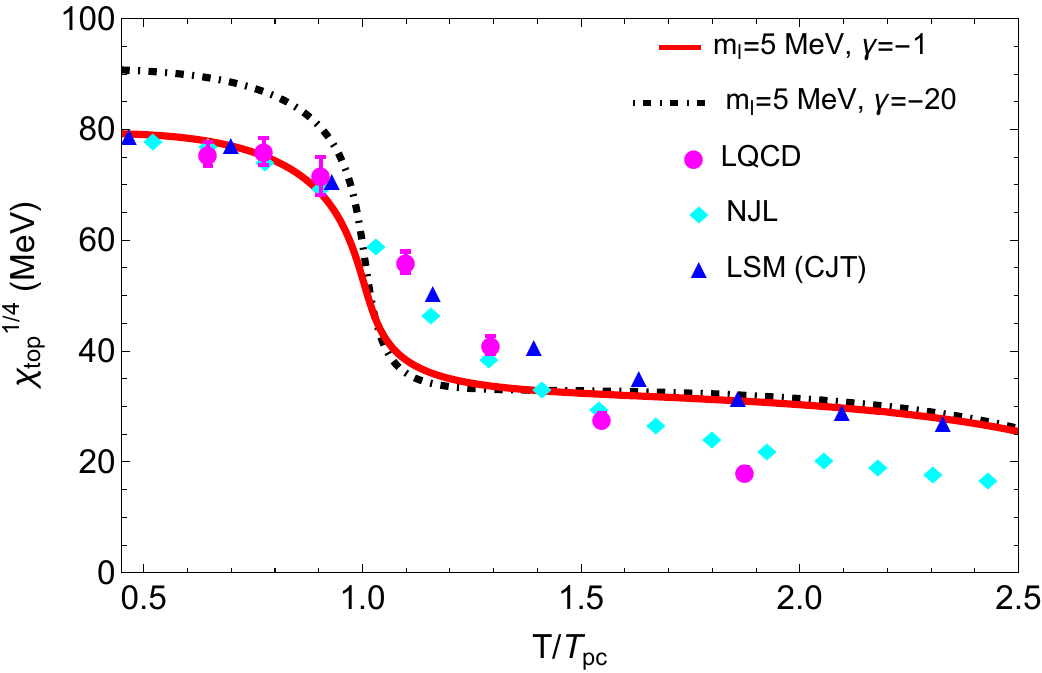}   
\caption{Left panel: The topological susceptibilities as a function of temperature at light quark masses $m_l=7.5$ MeV for different values of the determinant term's coupling as $\gamma=-1$ (solid red line) and $\gamma=-20$ (dot-dashed black line). Right panel: The color online is similar to the right panel for the light quark masses $m_l=5$ MeV. }
\label{topB}
\end{figure} 

The effect of $U(1)_A$ symmetry breaking was introduced through the determinant term in our model. The strength of the determinant term is controlled by the coupling parameter $\gamma$. To analyze the effect of the determinant term, we computed the topological susceptibility for different values of $\gamma$. As shown in Fig. \ref{topB}, for a fixed light quark mass ( $m_l=7.5$ MeV or $5$ MeV), a larger negative value of $\gamma$ simply shifts the entire $\chi_{\text{top}}^{1/4}$ curve to a higher overall magnitude. Conversely, a smaller magnitude yields a lower curve. Crucially, the functional shape of the curve remains unchanged. This result has a significant physical implication. It demonstrates that in our model, the determinant term acts primarily as a static background source for topological effects, scaling their overall intensity up or down. However, it fails to modify the underlying dynamical response of the topological sectors. The gravitational background, which dictates how the system evolves, still forces the topological susceptibility to be strongly tied to the chiral transition scale, regardless of the value of $\gamma$.

\section{Conclusions}
\label{sectionVI}

In this work, we analyzed the finite-temperature restoration of chiral and $U(1)_A$ symmetries within the framework of a soft-wall holographic QCD model. Two specific parameterizations, denoted as Case I and Case II, were employed, both tuned to yield a physical pion mass and a chiral pseudocritical temperature of approximately 155 MeV. The primary observables were the quark condensates, meson screening masses, meson susceptibilities, and the topological susceptibility.

Our calculations demonstrate that the model robustly predicts a smooth chiral crossover transition. The temperature evolution of the light-quark condensate $\sigma_l$ shows a smooth decrease, with its inflection point defining a pseudocritical temperature of $T_{\rm pc} = 0.157$ GeV for Case I and $T_{\rm pc} = 0.154$ GeV for Case II. This crossover behavior is a key feature of QCD with physical quark masses. The behavior of meson screening masses further confirms the restoration of chiral symmetry. The computed masses for the chiral partners ($\pi$, $\sigma$) and ($\eta$, $a_0$) become degenerate around $T_{\rm pc}$, a direct signal of the effective restoration of chiral symmetry. This finding is quantitatively consistent with results from lattice QCD and effective models, such as the NJL model.

The analysis of meson susceptibilities provides a complementary and rigorous test. The differences $\chi_{\pi} - \chi_{\sigma}$ and $\chi_{\eta_l} - \chi_{a_0}$, which serve as indicators for chiral symmetry restoration, approach zero sharply around $T_{\rm pc}$ for both model parameterizations. The associated pseudocritical temperatures extracted from these indicators align with those obtained from the quark condensate, supporting a consistent picture of chiral restoration.

A central finding of this study concerns the fate of the $U(1)_A$ symmetry. From the indicator $\chi_{\pi} - \chi_{a_0}$, which measures explicit $U(1)_A$ breaking due to the anomaly, we can see that the restoration of the $U(1)_A$ symmetry is later than the chiral symmetry. This result, which holds for both Case I and Case II, indicates that within our holographic setup, the restoration scales for chiral and $U(1)_A$ symmetries do not coincide, which suggests a possible separation between these scales. Despite the inconsistency of the indicator of $U(1)_A$ axial anomaly as a function of temperature with LQCD for temperatures $T<0.175$ GeV, the higher temperature region match together.

The gravitational dual and the incorporated determinant term, parameterized by $\gamma$, tie the dynamics of the axial anomaly to the chiral order parameter. While varying $\gamma$ alters the overall magnitude of anomaly-related quantities like the topological susceptibility, it does not decouple their temperature evolution from the chiral transition. This suggests that a more intricate holographic mechanism, perhaps involving independent bulk dynamics for topological sectors, is required to capture the potential persistence of $U(1)_A$ breaking effects above $T_{\rm pc}$.

The topological susceptibility $\chi_{\rm top}^{1/4}$, calculated via the Ward-Takahashi identity, exhibits a characteristic temperature dependence: it remains nearly constant for $T < T_{\rm pc}$, drops sharply around the transition, and then slowly decreases at higher temperatures. While this qualitative trend agrees with other approaches, the detailed comparison reveals that our model predicts a more pronounced drop near $T_{\rm pc}$ and a slower subsequent suppression. This behavior is a direct consequence of the effect of the chiral symmetry restoration on the $U(1)_A$ axial anomaly in our framework.

In summary, the soft-wall holographic model successfully describes key aspects of the chiral phase transition, including the crossover nature, the degeneracy of chiral partners, and the associated susceptibility patterns. Moreover, it provides a separate restoration scale for chiral and $U(1)_A$ symmetries. Future improvements of the model should aim to introduce more independent elements into the gravitational dual, allowing for a better description of the $U(1)_A$ anomaly for achieving a closer alignment with the emerging picture from first-principles lattice QCD studies.

\begin{acknowledgments} 
This work is supported in part by the National Natural Science Foundation of China (NSFC) Grant Nos: 12235016,
12221005, 12275108, and W2433019.
\end{acknowledgments}

\appendix
\section{ The value of mass parameters}
\label{appendA}

\begin{equation}
M_s^{a,b}(z) =\left(\begin{array}{cccc}
(\frac{3}{2} \chi_l^{2})_{I_{3 \times 3}} &   0   &  0  &   0 \\
   0   &  (\frac{1}{2} (\chi_l^{2} +\chi_l \chi_s +\chi_s^{2})  )_{I_{4 \times 4}} & 0 & 0 \\
   0   &  0 &  \frac{1}{2} (\chi_l^{2} +2 \chi_s^{2}) & \frac{1}{\sqrt{2}} (\chi_l^{2} - \chi_s^{2}) \\
   0   &  0 & \frac{1}{\sqrt{2}} (\chi_l^{2} - \chi_s^{2}) &  \frac{1}{2} (2\chi_l^{2} + \chi_s^{2})
\end{array}\right),
\end{equation}

\begin{equation}
M_{det}^{a,b}(z) =\left(\begin{array}{cccc}
(\frac{-1}{8} \chi_s)_{I_{3 \times 3}} &   0   &  0  &   0 \\
   0   &  (\frac{-1}{8} \chi_l )_{I_{4 \times 4}} & 0 & 0 \\
   0   &  0 &  \frac{-1}{24} (4\chi_l - \chi_s) & \frac{-1}{12 \sqrt{2}} (\chi_l - \chi_s) \\
   0   &  0 & \frac{-1}{12 \sqrt{2}} (\chi_l - \chi_s) &  \frac{1}{12} (2\chi_l + \chi_s)
\end{array}\right),
\end{equation}

\begin{equation}
M_A^{a,b}(z) =\left(\begin{array}{cccc}
( \chi_l^{2})_{I_{3 \times 3}} &   0   &  0  &   0 \\
   0   &  (\frac{1}{4} (\chi_l +\chi_s)^2  )_{I_{4 \times 4}} & 0 & 0 \\
   0   &  0 &  \frac{1}{3} (\chi_l^{2} +2 \chi_s^{2}) & \frac{\sqrt{2}}{3} (\chi_l^{2} - \chi_s^{2}) \\
   0   &  0 & \frac{\sqrt{2}}{3} (\chi_l^{2} - \chi_s^{2}) &  \frac{1}{3} (2\chi_l^{2} + \chi_s^{2})
\end{array}\right),
\end{equation}
where the index $a,b=9$ corresponds to the singlet state.

%

 \addcontentsline{toc}{section}{References}

\end{document}